\magnification=\magstephalf
\input amstex
\loadbold
\documentstyle{amsppt}
\refstyle{A}
\NoBlackBoxes

\vsize=7.5in

\def\pf{\hfill $\square$}
\def\c{\cite}

\def\fg{\frak{g}}
\def\fh{\frak{h}}

\def\end{\text{End}}

\def\bla{\boldsymbol \lambda}
\def\bn{\boldsymbol \nu}

\def\bbl{\boldkey l}

\def\bA{{\bold A}{\bold d}}

\topmatter
\title A class of integrable spin Calogero-Moser systems II: exact 
solvability
  \endtitle
\leftheadtext{L.-C. Li}
\rightheadtext{exact solvability of spin CM systems}
\author Luen-Chau Li\endauthor
\address{L.-C. Li, Department of Mathematics,Pennsylvania State University,
University Park, PA  16802, USA}\endaddress
\email luenli\@math.psu.edu\endemail
\smallskip
\dedicatory{Dedicated with respect and admiration to Percy Deift
on the occasion of his 60th birthday}\enddedicatory
\abstract  In a previous paper, we introduce a class of integrable spin 
Calogero-Moser systems associated with the classical dynamical r-matrices 
with spectral parameter.  Here the main purpose is to give explicit
solutions of several factorization problems associated with infinite 
dimensional Lie groupoids which will allow us to write down the solutions of
these integrable models.

\endabstract
\endtopmatter

\document
\subhead
1. \ Introduction
\endsubhead

\baselineskip 15pt
\bigskip

In \c{LX1},\c{LX2}, we introduce a class of spin Calogero-Moser (CM) systems 
associated with so-called classical dynamical r-matrices with spectral 
parameter, as defined and classified in \c{EV} for simple Lie algebras,
following the pioneering work of Felder \c{F} and Felder and
Wieczerkowski \c{FW} in which
the classical dynamical Yang-Baxter equation (CDYBE) with spectral parameter
was introduced and studied in the context of conformal field
theory.  The main 
purpose of this sequel is to show how to obtain the explicit solutions 
of the associated integrable systems in 
\c{LX2} by using the factorization method developed in \c{L2}.

The spin CM systems constructed in the afore-mentioned papers are of three 
types-rational, trigonometric and elliptic.  Indeed, for each of
the canonical forms of the three types of $z$-dependent classical 
dynamical r-matrices in \c{EV}, there is an intrinsic way to
construct an associated spin CM system and its realization
in the dual bundle of a corresponding coboundary dynamical Lie
algebroid.  In this way, we are led to a family of rational spin CM
systems parametrized by subsets $\Delta^{\prime}\subset \Delta$ which
are closed with respect to addition and multiplication by $-1$.  Here
$\Delta$ is the root system associated with a complex simple Lie
algebra $\fg$ with a fixed Cartan subalgebra $\fh.$ In the
trigonometric case, we also have a family but here the family is 
parametrized by
subsets $\pi^{\prime}$ of a fixed simple system $\pi\subset \Delta.$ 
Finally, we have an elliptic spin Calogero-Moser system for
every simple Lie algebra.  Let us summarize a few key features of
these Hamiltonian systems and their realization spaces as follows: 
(a) in each case the phase space $P$ is a Hamiltonian $H$-space 
(with equivariant momentum map $J$) which admits an $H$-equivariant 
realization in the dual bundle $A\Gamma$ of an infinite dimensional 
coboundary dynamical Lie algebroid $A^{*}\Gamma$ and the Hamiltonian is 
the pullback of a natural invariant function on $A\Gamma$ under the 
realization map $\rho$, 
(b) the coboundary dynamical Lie algebroids involved are associated with  
solutions of the modified dynamical Yang-Baxter equation (mDYBE),
(c) the pullbacks of the natural invariant functions on $A\Gamma$ by $\rho$
do not Poisson commute everywhere on $P$, but they do so on the fiber
$J^{-1}(0)$ in all cases,  
(d) the reduced Hamiltonian system on $J^{-1}(0)/H$ admits a natural
collection of Poisson commuting integrals.  As a matter of fact,
we now know from \c{L2} that  $A\Gamma$ is also a Hamiltonian
$H$-space  and it follows from the same work that the integrable
flows on the reduced space are  realized on the Poisson
quotient $\gamma^{-1}(0)/H$, where 
$\gamma: A\Gamma\longrightarrow \fh^*,\,\, (q,\lambda,X)\mapsto \lambda$
is the momentum map of the $H$-action on $A\Gamma.$   Indeed we have 
$\rho(J^{-1}(0))\subset \gamma^{-1}(0)$ in each case.  Since $\rho$
is $H$-equivariant, it therefore induces a Poisson map between
the corresponding Poisson quotients.

We now turn to our approach on exact solvability. As we showed in 
\c{L2}, the mDYBE is associated with factorization problems on 
trivial Lie groupoids the solutions of which provide an effective
method to integrate the generalized Lax equations on $\gamma^{-1}(0).$
(See \c{L1} for the groupoid version.)  In the context of our
spin CM systems, the factors which appear in the factorization problems are 
elements of certain Lie subgroupoids of trivial Lie groupoids of 
the form $\Gamma=U\times LG\times U,$ 
where $LG$ is the loop group associated with a simple Lie group $G$, and
$U$ is an open subset of a Cartan subalgebra $\fh\subset\fg= Lie(G).$
Consequently, it is essential to have a precise description of
the elements which belong to these Lie subgroupoids.
As a first step towards the solutions of the factorization
problems, we begin by analyzing the corresponding
decompositions on the infinitesimal level, i.e., at the level
of Lie algebroids.  Here, the $r$-matrix
${\Cal R}:A^{*}\Gamma\longrightarrow A\Gamma$ of the 
coboundary dynamical Lie algebroid $A^{*}\Gamma$ plays
an important role.  Indeed, for associated bundle maps
${\Cal R}^{\pm}:A^{*}\Gamma\longrightarrow A\Gamma$,
$Im\,{\Cal R}^{\pm}$
are Lie subalgebroids of the trivial Lie algebroid
$A\Gamma$ and the infinitesimal version of factorization
is the decomposition
$$(0_q, X,0) = {1\over 2}{\Cal R}^{+}(0_q,X,0)
             - {1\over 2}{\Cal R}^{-}(0_q,X,0)$$
where the element $(0_q,X,0)$ on the left hand side
of the above expression is in the adjoint bundle
of $A\Gamma.$  Thus it is essential to be able to describe 
the elements of  $Im\,{\Cal R}^{\pm}.$  As it turns out,
$Im\,{\Cal R}^{+} = \bigcup_{q\in U} \{0_q\} \times L^{+}\fg \times \fh$
in all three cases, where $L^{+}\fg$ is the Lie subalgebra of
the loop algebra $L\fg$ consisting of convergent
power series $\sum_{0}^{\infty} X_{n}z^{n}.$
On the other hand, $Im\,{\Cal R}^{-}$  in each of the three cases is given by
a matched product ${\Cal I}^{-}\bowtie{\Cal Q}$ (in the sense of
Mokri \c{Mok}) where
the ideal ${\Cal I}^{-}$ coincides with the adjoint
bundle of  $Im\,{\Cal R}^{-}$ and ${\Cal Q}$ is a Lie
subalgebroid of  $Im\,{\Cal R}^{-}$ isomorphic to $TU.$
In spite of this, the method of solution of the 
factorization problems is quite different in the three cases under
consideration.  That this is so is due to the  difference in the 
analyticity properties
of the elements in the ideals ${\Cal I}^{-}$.
In the rational case and trigonometric case, we can reduce the
factorization problems to finite dimensional problems 
due to some special features of the Lax operators. (Of course,
this is also a reflection of the underlying character of
the flows.) However, this is
not so in the elliptic case-here we will only do things 
for the classical Lie algebras and indeed we will only give
details for $\fg= sl(N,\Bbb{C})$ as the arguments for
the other classical Lie algebras are similar.  In this case,
the explicit solution of the factorization problem is obtained with
the help of multi-point Baker-Akheizer functions connected with the 
spectral curve $C$.  Thus the solution of the equations 
can be expressed in terms of Riemann theta functions.

The paper is organized as follows.  In Section 2, we present a number
of basic results which will be used throughout the paper.  More
specifically, we will begin with a review on the geometric scheme to 
construct integrable systems based on realization in the dual bundles
of coboundary dynamical Lie algebroids and the factorization
theory which we mentioned earlier.  Then we will turn our attention
to a subclass of coboundary dynamical Lie algebroids defined by
the classical dynamical r-matrices with spectral parameter.
We will also recall what we mean by spin Calogero-Moser systems
associated with this subclass of coboundary dynamical Lie algebroids.  
In Section 3, we discuss the solution of the integrable rational spin 
Calogero-Moser systems by solving the corresponding factorization
problem.  In Section 4, we handle the trigonometric case.
Finally in Section 5, we analyze the elliptic case.  

To close, we would like to point out what was done
in the paper \c{KBBT} so that the reader can better understand
why a different method is required for our more general
class of systems here.
To cut the story short, what the authors considered
in \c{KBBT} are the $gl(N,\Bbb C)$-rational spin Calogero-Moser
system of Gibbons and Hermsen \c{GH}, as well as their
trigonometric and elliptic counterparts.  From our
point of view, these are special cases which can be
obtained from more general 
$gl(N,\Bbb C)$-systems (see \c{BAB} and \c{ABB})
by restricting the matrix of `spin variables' 
to some {\it special\/} coadjoint orbits of 
$gl(N,\Bbb C)^*\simeq gl(N,\Bbb C)$ which can be parametrized
by vectors $a_j,b_j\in \Bbb C^{l}$, $l<N,\,j=1,\cdots, N.$
The method for solving such systems in \c{KBBT} is based
on the connection with the matrix KP equation and is
specific to these special coadjoint orbits of
$gl(N,\Bbb C)^*\simeq gl(N,\Bbb C)$.  
For a sketch of this method in the elliptic case
together with an explanation of its limitations,
we refer the reader to Remark 5.2.8 (a).

\bigskip
\bigskip

\subhead
2. \ Invariant Hamiltonian systems associated with coboundary dynami-
\linebreak \phantom{fak} cal Lie algebroids and the factorization method
\endsubhead

\bigskip

In the first two subsections, we shall present a number of basic results from 
\c{L2} which will be used throughout the paper.  In particular, we shall
give a summary of the factorization theory.  In the last subsection, we
shall turn our attention to a subclass of coboundary dynamical Lie
algebroids defined by the classical dynamical r-matrices with
spectral parameter \c{LX2}. Since the paper is concerned
with the solution of the class of integrable spin Calogero-Moser systems 
introduced in \c{LX2}, we will recall its construction and its
relation to this subclass of Lie algebroids in this last subsection.
\smallskip

\subhead
2.1 \ Geometry of the  modified dynamical Yang-Baxter equation 
\endsubhead
\medskip

Let $G$ be a connected Lie group, $H\subset G$ a connected Lie subgroup, 
and $\fg$, $\fh$ their Lie algebras. We shall denote by  
$\iota :\fh \longrightarrow \fg$ the Lie inclusion.  In what follows,
the Lie groups and Lie algebras can be real or complex unless we
specify otherwise.

If $U\subset \fh^*$ is a connected $Ad_H^*$-invariant open subset,
we say that a smooth (resp.~ holomorphic) map  
$R:U\longrightarrow L(\fg^*, \fg)$ 
(here and henceforth we denote by
$L(\fg^*,\fg)$ the set of linear maps from $\fg^*$ to $\fg$)
is a classical dynamical r-matrix \c{EV} associated with the
pair $(\fg,\fh)$ iff $R$ is
pointwise  skew symmetric
$$<R(q)(A), B>=- <A, R (q) B>\eqno (2.1.1)$$
and satisfies the classical dynamical Yang-Baxter condition
$$\eqalign {&[R(q)A, R(q)B] +R(q)(ad^*_{R(q)A}B-ad^*_{R(q)B}A)\cr
+&dR(q)\iota^*A(B) - dR(q)\iota^*B(A) + d<R(A),B>(q) 
= \chi (A,B),\cr} \eqno (2.1.2)$$
for all $q\in U$, and all $A, B\in \fg^*$, 
where $\chi : \fg^*\times \fg^* \longrightarrow  \fg$ is
$G$-equivariant.
The dynamical $r$-matrix is said to be $H$-equivariant
if and only if
$$R(Ad^*_{h^{-1}} q)=Ad_h \circ R(q)\circ Ad^*_h \eqno (2.1.3)$$
for all $h\in H, q\in U.$  

We shall equip 
$\Gamma=U\times G\times U$ with the trivial Lie groupoid
structure over $U$ with structure maps (target, source,\,$\ldots,$
multiplication)
$$\eqalign {&\alpha (u,g,v)= u,\, \beta (u,g,v)=v,\, \epsilon (u) = (u,
1,u), \, 
i(u,g,v)= (u, g^{-1}, v)\cr
&\hskip 60 pt m ((u,g,v), (v,g^{\prime}, w))= (u, gg^{\prime}, w)\cr}
\eqno (2.1.4)$$
and let
$A \Gamma = Ker\, T\alpha|_{\epsilon(U)} =\bigcup_{q\in U} \lbrace 0_{q} \rbrace
\times \fg \times \fh^* \simeq TU\times \fg$ be its Lie 
algebroid with anchor map denoted by $a.$
(See \c{CdSW} and \c{M1} for details.)  Recall that
associated with an $H$-equivariant classical
dynamical r-matrix $R$ there is a natural Lie algebroid bracket 
$[\cdot, \cdot]_{A^*\Gamma}$ on the dual bundle $A^*\Gamma$ of 
$A\Gamma$ \c{BKS},\c{L2} such that  the pair
$(A\Gamma, A^*\Gamma)$ is a Lie bialgebroid in the sense of MacKenzie
and Xu \c{MX}.(Lie bialgebroids are infinitesimal versions of the
Poisson groupoids of Weinstein \c{W}.)  Throughout the paper, the pair
$(A^{*}\Gamma, [\cdot,\cdot]_{A^{*}\Gamma})$
together with the anchor map $a_{*}:A^{*}\Gamma\longrightarrow TU$
given by
$$a_{*} (0_q,A,Z)=(q, \iota^{*}A-ad^{*}_{Z}q)\eqno(2.1.5)$$
will be called the 
{\it coboundary dynamical Lie algebroid\/} associated to $R$.

A special case of (2.1.2) is the modified dynamical Yang-Baxter equation 
(mDYBE):
$$\eqalign {& [R(q)A, R(q)B]+R(q)(ad^*_{R(q)A}B-ad^*_{R(q)B}A)\cr
+\,&dR(q)\iota^*A(B) - dR(q)\iota^*B(A) + d<R(A),B>(q)\cr
=\,& - [K(A), K(B)]\cr}\eqno(2.1.6)$$
where $K\in L(\fg^*,\fg)$ is a nonzero symmetric map which
satisfies $ad_X\circ K + K\circ ad^*_X = 0$ for all $X \in \fg$,i.e.,
$K$ is $G$-equivariant.  In \c{L2}, the class of coboundary dynamical
Lie algebroids associated with mDYBE was singled out and was shown
to have some rather remarkable properties.
We will restrict to this class of 
$\left(A^{*}\Gamma, [\cdot,\cdot]_{A^{*}\Gamma}, a_{*}\right)$ in the
rest of the subsection.

Following \c{L1} and \c{L2}, we introduce the bundle map  
$${\Cal R}: A^*\Gamma \longrightarrow A\Gamma, 
(0_{q}, A, Z)\mapsto (0_{q}, -\iota Z+R(q)A, \iota^*A-ad^*_{Z}q)
\eqno(2.1.7)$$ 
and call it the {\sl r-matrix of the Lie algebroid} $A^*\Gamma$.
Also, define
$${\Cal K}:A^*\Gamma \longrightarrow A\Gamma, 
(0_q,A,Z) \mapsto (0_q,K(A),0),\eqno(2.1.8)$$
and set 
$${\Cal R}^{\pm}={\Cal R}\pm {\Cal K},\,\, R^{\pm}(q)=R(q)\pm K.\eqno(2.1.9)$$
\proclaim
{Proposition 2.1.1} (a) ${\Cal R}^{\pm}$ are morphisms of transitive
Lie algebroids and, as morphisms of vector bundles over $U$, are of
locally constant rank.   
\smallskip
\noindent (b) $Im {\Cal R}^{\pm}$ are transitive Lie subalgebroids of 
$A\Gamma.$
\endproclaim

In the rest of the subsection,we shall assume $\fg$ has a nondegenerate 
invariant pairing $(\cdot, \cdot)$ such that 
$(\cdot,\cdot)\mid_{\fh\times \fh}$ is also nondengenerate.
Without loss of generality,
we shall take the map $K:\fg^* \longrightarrow \fg$ in the above
discussion to be the identification map induced by $(\cdot, \cdot)$.
Indeed, with the identifications $\fg^* \simeq \fg$, 
$\fh^*\simeq \fh$, we have $K = id_{\fg}.$ We shall regard
$R(q)$ as taking values in $End (\fg)$, and the derivatives
as well as the dual maps are computed using $(\cdot, \cdot)$.  Also,
we have $ad^* \simeq -ad$, $\iota^* \simeq \Pi_{\fh}$, where $\Pi_{\fh}$
is the projection map to $\fh$ relative to the direct sum decomposition
$\fg = \fh \oplus \fh^{\perp}$.  We shall keep, however, the notation
$A^*\Gamma$ although as a set it can be identified with $A\Gamma.$

We now introduce the following subbundles of the adjoint bundle
$Ker\,a = \bigl \{ (0_q,X,0)\mid q\in U, X\in \fg \bigr \}$ of \, $A\Gamma$:
$${\Cal I}^{\pm} =\bigl \{(0_q,X,0)\in Ker\,a\mid q\in U,
   {\Cal R}^{\mp} (0_q,X,Z)=0 \,\,\,\hbox{for some}\,\,\, Z \in \fh \bigr \}.
\eqno(2.1.10)$$

\proclaim
{Proposition 2.1.2} (a) ${\Cal I}^{\pm}$ are ideals of the transitive
Lie algebroids $Im {\Cal R}^{\pm}$.
\smallskip
\noindent (b) Equip $Im {\Cal R}^{+}/{\Cal I}^{+}$ and 
$Im {\Cal R}^{-}/{\Cal I}^{-}$ with the quotient transitive
Lie algebroid structures, then
the map $\theta : Im {\Cal R}^{+}/{\Cal I}^{+}
\longrightarrow Im {\Cal R}^{-}/{\Cal I}^{-}$ defined by
$$\theta ({\Cal R}^{+} (0_q,X,Z) + {\Cal I}^{+}_q)=
  {\Cal R}^{-} (0_q,X,Z) + {\Cal I}^{-}_q$$
is an isomorphism of transitive Lie algebroids.
\endproclaim

\proclaim
{Theorem 2.1.3} (a) Every element $(0_q,X,0)\in Ker\,a$ admits a
unique decomposition
$$(0_q,X,0) = {\Cal X}_+ - {\Cal X}_-$$
where $({\Cal X}_+ ,{\Cal X}_-) \in ({\Cal R}^+, {\Cal R}^-)(\{0_q\}
\times \fg \times \{0\})$ with 
$\theta({\Cal X}_+ + {\Cal I}^{+}_q) = {\Cal X}_{-} + {\Cal I}^{-}_q.$
\smallskip
\noindent (b) The coboundary dynamical Lie algebroid $A^*\Gamma$ is
isomorphic to the Lie subalgebroid
$$ \left \{({\Cal X}_+,{\Cal X}_-)\in (Im{\Cal R}^{+}
      {\underset TU \to\oplus}Im{\Cal R}^{-})_{q}\mid q\in U,\, 
      \theta({\Cal X}_{+}+{\Cal I}^{+}_q) = 
      {\Cal X}_{-} + {\Cal I}^{-}_q \right \}$$
of  $Im{\Cal R}^{+} {\underset TU \to\oplus}Im{\Cal R}^{-}$.
\endproclaim
\medskip

\subhead
2.2 \ Invariant Hamiltonian systems and the factorization method
\endsubhead
\medskip

We shall continue to use the same assumptions on $\fg$, $K$ and $R$
as in the latter part of Section 2.1. Although it is not entirely
necessary to assume that $R$ satisfies the mDYBE for Theorem 2.2.1,
however, it serves our purpose here.

Let $P$ be a Poisson manifold and suppose
$\rho = (m, \tau, L): P \longrightarrow U\times \fh \times \fg \simeq
A\Gamma$ is a realization of  $P$ in the dual bundle
$A\Gamma$ of the coboundary dynamical Lie algebroid $A^{*}\Gamma$,
equipped with the Lie-Poisson structure, i.e., 
$\rho$ is a Poisson map.  Recall that $A\Gamma$ is a Hamiltonian
$H$-space under the natural action
$h.(q,\lambda, X) = (Ad^{*}_{h^{-1}}q, Ad^{*}_{h^{-1}}\lambda, Ad_{h}X)$
and the projection map
$\gamma: A\Gamma\longrightarrow \fh^*,\,\, (q,\lambda,X)\mapsto \lambda$
is an equivariant momentum map.
We assume:
\medskip
\noindent A1. $P$ is a Hamiltonian $H$-space with an equivariant
momentum map $J:P\longrightarrow \fh$,\newline
\noindent A2. the realization map $\rho$ is $H$-equivariant,\newline
\noindent A3. for some regular value $\mu\in \fh$ of $J$, 
$$\rho (J^{-1} (\mu)) \subset \gamma^{-1}(0)=U\times \{0\}\times \fg. 
\eqno(2.2.1)$$

Let $I(\fg)$ be the ring of smooth ad-invariant functions on $\fg$, 
$i_{\mu}: J^{-1}(\mu)\longrightarrow P$  the inclusion map, and
$\pi_{\mu}: J^{-1}(\mu)\longrightarrow J^{-1}(\mu)/H_{\mu}$ the canonical
projection, where $H_{\mu}$ is the isotropy subgroup of $\mu$
for the $H$-action on $P.$ Also, let $Pr_{3}$ denote the projection
map onto the third factor of $U\times \fh\times \fg\simeq A\Gamma.$
We consider $H$-invariant Hamiltonian systems on $P$, generated by Hamiltonians
of the form ${\Cal F}= L^{*}f$, where $f \in I(\fg)$. 
 
\proclaim
{Theorem 2.2.1} Under the above assumptions, 
\smallskip
\noindent (a) The Hamiltonian ${\Cal F}$ descends to a unique function
${\Cal F}_{\mu}$ on the reduced Poisson variety $P_{\mu} = J^{-1}(\mu)/H_{\mu}.$
Moreover, ${\Cal F}_{\mu}$ admits a natural family of Poisson commuting 
functions given by the reduction of functions in $L^{*} I(\fg)$,
\smallskip
\noindent (b) If $\psi_t$ is the induced flow on 
$\gamma^{-1}(0)$ 
generated by the Hamiltonian $Pr^*_{3} f$, and 
$\phi_t$ is the Hamiltonian flow of ${\Cal F}$ on $P$, then
under the flow $\phi_{t}$, we have
$$
\aligned
       & {d\over dt} m(\phi_{t}) = \Pi_{\fh}
         df(L(\phi_{t})),\\
       & {d\over dt} \tau(\phi_{t}) = 0, \\
       & {d\over dt} L(\phi_{t}) = [\,L(\phi_{t}), R(m(\phi_{t}))
          df(L(\phi_{t}))\,] + dR(m(\phi_{t}))(\tau(\phi_{t}))
          df(L(\phi_{t}))
\endaligned
$$
where the term involving $dR$ drops out on $J^{-1}(\mu)$.
Moreover, the  reduction $\phi^{red}_t$ of $\phi_{t} \circ i_{\mu}$ on
$P_{\mu}$ defined by $\phi^{red}_t \circ \pi_{\mu} = \pi_{\mu} \circ
\phi_{t} \circ i_{\mu}$ is a Hamiltonian flow of ${\Cal F}_{\mu}.$
\endproclaim

This theorem applies in particular to the special case where $P= A\Gamma$, 
$\rho = id_{A\Gamma}$ (see \c{L2} for the proof that $A\Gamma$ is a
Hamiltonian $H$-space).  In fact, we have a factorization
method for solving the generalized Lax equations
$$\eqalign{& {d \over dt}\, (q,0,X) \cr
  =\, & (\,\Pi_{\fh}\, df(X), 0, [X, R(q)df(X)]\,)\cr}\eqno(2.2.2)$$
on the invariant manifold $\gamma^{-1}(0).$  
In what follows, if  $(u,g,v)\in \Gamma,$ the symbol 
$\bbl_{(u,g,v)}$  will stand for left translation
in $\Gamma$ by $(u,g,v)$, the lift of 
$Im({\Cal R}^+, {\Cal R}^-) \subset A\Gamma {\underset TU \to \oplus} A\Gamma$
to the groupoid level will be denoted by the same symbol, and
${\bA}$ is the adjoint representation of $\Gamma$ on its adjoint bundle
$Ker\,a$,  defined by ${\bA}_{\gamma} (q,0,X) = (q^{\prime},0,
Ad_{k}X)$, for $\gamma = (q^{\prime},k,q)\in \Gamma$.
Lastly, if $Y$ is a section of $A\Gamma$, the exponential
$exp\,(tY):U\longrightarrow \Gamma$ is defined by the
formula $exp(tY)(q) = f_{t}(\epsilon(q))$, where
$f_{t}$ is the local flow generated by the left-invariant
vector field $\overleftarrow Y$:
$\overleftarrow Y(u,g,v) = T_{\epsilon(v)} \bbl_{(u,g,v)} Y(v).$
In particular, if we take $Y$ to be the constant
section $(0,0,\xi)$ of $A\Gamma,$ $\xi\in \fg$,
an easy computation shows that
$exp\{t(0,0,\xi)\}(q) = (q, e^{t\xi},q).$

\proclaim
{Theorem 2.2.2} Suppose that $f\in I(\fg)$, $F=Pr^{*}_{3} f$ and $q_0\in U$,
where $U$ is simply connected.  Then for some 
$0 < T \leq \infty$, there exists a unique element 
$(\gamma_{+}(t),\gamma_{-}(t))=
((q_0, k_{+} (t), q(t)), (q_0, k_{-} (t), q(t))) \in
Im ({\Cal R}^{+}, {\Cal R}^{-})\subset \Gamma
{\underset U \times U \to \times}\Gamma$
for $0 \leq t <T$ which is smooth in t, solves the factorization
problem

$$exp \{2\, t(0,0,df(X_0))\}(q_0)
      =\,\gamma_{+} (t)\,\gamma_{-} (t)^{-1}\eqno(2.2.3)
$$
and satisfies 
$$\left(T_{\gamma_{+} (t)} {\bbl}_{{\gamma_{+}(t)}^{-1}} {\dot\gamma_{+}(t)},
T_{\gamma_{-} (t)}{\bbl}_{{\gamma_{-}(t)}^{-1}}{\dot \gamma_{-}(t)}\right)
\in \, 
({\Cal R}^+,{\Cal R}^-) (\{q(t)\}\times\{0\}\times\fg)
\eqno(2.2.4a)$$
with
$$\gamma_{\pm}(0) = (q_{0},1,q_{0}).\eqno(2.2.4b)$$

\noindent Moreover, the solution of Eqn.(2.2.2) with initial data
$(q,0,X)(0)= (q_0,0,X_0)$ is given by the formula
$$(q(t),0,X(t)) ={\bA}_{{\gamma_{\pm}(t)^{-1}}} (q_0,0,X_0).\eqno(2.2.5)$$
\endproclaim

\proclaim
{Corollary 2.2.3} Let $\psi_{t}$ be the induced flow on  
$\gamma^{-1}(0)$ as
defined in (2.2.5) and let $\phi_{t}$ be the Hamiltonian flow of 
${\Cal F} =L^{*} f$ on $P$, where $L = Pr_{3}\circ\rho$ for a realization
map $\rho:P \longrightarrow A\Gamma$ satisfying A1-A3.  If we can solve
for $\phi_{t} (x)$, $x \in J^{-1} (\mu)$ explicitly from the relation
$\rho(\phi_{t}) (x) = \psi_{t} (\rho (x))$, then the formula
$\phi^{red}_{t} \circ \pi_{\mu} = \pi_{\mu} \circ \phi_{t} \circ i_{\mu}$
gives an explicit expression for the flow of the reduced Hamiltonian
${\Cal F}_{\mu}$.
\endproclaim
\medskip

\subhead
2.3 \ Classical dynamical r-matrices with spectral parameter and
the \linebreak \phantom{fake}\,\,\, associated spin Calogero-Moser systems
\endsubhead
\medskip

From now onwards, we let $\fg$ be a complex simple Lie algebra of rank
$N$ with Killing form $(\cdot,\cdot)$ and let $G$ be the connected
and simply-connected Lie group which integrates $\fg.$
We fix a Cartan subalgebra $\fh$ of $\fg$ and let
 $\fg=\fh \oplus \sum_{\alpha \in\Delta} \fg_{\alpha}$ be
the root space decomposition of $\fg$ with respect to $\fh$. 
For each $\alpha \in \Delta$, denote by $H_{\alpha}$ the element
in $\fh$ which corresponds to $\alpha$ under the isomorphism
between $\fh$ and $\fh^*$ induced by the Killing form $(\cdot,\cdot)$.
We fix a simple system of roots  $\pi=\{\alpha_1,\cdots, \alpha_N\}$ 
and denote by $\Delta^\pm$ the corresponding positive/negative
system. Also, for each $\alpha\in \Delta^+$, we pick root vectors
$e_{\alpha}\in \fg_{\alpha}$, $e_{-\alpha}\in \fg_{-\alpha}$
which are dual with respect to $(\cdot,\cdot)$ so
that $[e_{\alpha},e_{-\alpha}] = H_{\alpha}.$ 

Let $r :\fh\times \Bbb C\longrightarrow \fg\otimes\fg$ 
be a classical dynamical r-matrix with spectral parameter
in the sense of \c{EV}, with coupling constant equal to $1$.
Then we can construct the associated classical dynamical r-matrix
$R$  \c{LX2} for the pair $(L\fg,\fh)$, where $L\fg$ is the loop
algebra of $\fg$.
Indeed, if $L\fg^*$ denotes the restricted dual of $L\fg$ and we
make the identification $L\fg^*\simeq L\fg$, then we have
the following result.

\proclaim
{Proposition 2.3.1} For $X\in L\fg$, we have the formula
$$(R(q)X )(z)= {1\over 2} X(z) +
\sum_{k\geq 0}\frac{1}{k!}\left ( \frac{\partial^{k}r}{\partial z^{k}}
(q, -z ), \ X_{-(k+1)}\otimes 1 \right ). \eqno(2.3.1)$$
Moreover, $R$ is a solution of the mDYBE with $K = {1\over 2} id_{L\fg}$.
\endproclaim
\smallskip
\noindent {\bf Remark 2.3.2.} As the reader will see, the formula
in (2.3.1) will be used to compute the explicit expressions for $R$ 
which play a critical role in characterizing the elements in the Lie 
subalgebroids $Im {\Cal R}^{\pm}$ in our analysis in Section 3, 4 and 5 below.
This formula has also been used in connection with symmetric
coboundary dynamical Lie algebroids in \c{L3}.
\smallskip

We now fix a simply-connected domain $U\subset \fh$ on which
$R$ is holomorphic.  Also, introduce the trivial Lie
groupoids $\Omega = U\times G\times U$, $\Gamma = U\times LG\times U,$
where $LG$ is the loop group of the simple Lie group $G.$
Then we can use the map $R:U\longrightarrow L(L\fg, L\fg)$
in Proposition 2.3.1 to construct the associated
coboundary dynamical Lie algebroid
$A^{*}\Gamma = T^{*} U \times L\fg^* \simeq TU \times L\fg$
so that its dual bundle has a Lie-Poisson structure.
(See \c{L2} for explicit formulas.)  On the other
hand, we shall equip the dual bundle $A^{*}\Omega\simeq TU\times \fg$
of the trivial Lie algebroid $A\Omega$ with the
corresponding Lie-Poisson structure.  Explicitly,
$\{\varphi,\psi\}_{A^{*}\Omega}(q,p,\xi) =(\delta_{2}\varphi,\delta_{1}\psi)
-(\delta_{1}\varphi, \delta_{2}\psi)+ (\xi,[\delta\varphi,\delta\psi])$
where $\delta_{1}\varphi,$ $\delta_{2}\varphi$ and
$\delta\varphi$ are the partial derivatives of 
$\varphi$ with respect to the variables in $U$, $\fh$ and
$\fg$ respectively.

\proclaim
{Theorem 2.3.3}  The map 
$\rho = (m,\tau, L): A^{*}\Omega\simeq TU \times \fg \longrightarrow 
TU\times L\fg \simeq A\Gamma$ 
given by
$$\rho (q,p,\xi) = (q, -\Pi_{\fh}\xi, p + r^{\#}_{-}(q)\xi)\eqno(2.3.2)$$
is an H-equivariant Poisson map, when the domain is equipped
with the Lie-Poisson structure corresponding to the trivial Lie
algebroid $A\Omega\simeq TU\times \fg$, and the target is equipped with
the Lie-Poisson structure corresponding to $A^*\Gamma$.  Here,
the map $r^{\#}_{-} (q) : \fg \longrightarrow L\fg$ is defined by
$$((r^{\#}_{-} (q)\xi)(z), \eta) = (r(q,z), \eta \otimes \xi) \eqno(2.3.3)$$
for $\xi$, $\eta \in \fg.$
\endproclaim
\smallskip

Let $Q$ be the quadratic function
$$Q(X)= {1\over 2}\oint_c (X(z), X(z))\frac{dz}{2\pi iz}\eqno(2.3.4)$$
where $c$ is a small circle around the origin.  Clearly, 
$Q$ is an ad-invariant function on $L\fg$.

\definition
{Definition 2.3.4}  Let $r$ be a classical dynamical r-matrix with
spectral parameter with coupling constant equal to 1.  Then the
Hamiltonian system on $A^{*}\Omega\simeq TU\times \fg$ 
(equipped with the Lie-Poisson
structure as in Theorem 2.3.3) generated by the Hamiltonian
$${\Cal H}(q,p,\xi)= (L^{*}Q)(q,p,\xi)= 
{1\over 2}\oint_c (L(q, p, \xi)(z), L(q, p, \xi)(z))\frac{dz}{2\pi iz}
\eqno(2.3.5)$$
is called the spin Calogero-Moser system associated to $r$.
\enddefinition
\smallskip

\proclaim
{Proposition 2.3.5} The Hamiltonians of the spin Calogero-Moser systems
are invariant under the Hamiltonian $H$-action on 
$A^{*}\Omega\simeq TU\times \fg$ given
by $h\cdot(q,p,\xi)=(q,p,Ad_{h} \xi)$ with equivariant momentum map 
$J(q,p, \xi) = - \Pi_{\fh}\, \xi.$  Moreover, under the Hamiltonian flow,
we have 
$$\eqalign{& \dot q  = \Pi_{\fh} (M(q,p,\xi))_{-1}, \cr
           & \dot L (q,p,\xi) = [\, L(q,p,\xi), R(q) M(q,p,\xi) \,]}
           \eqno(2.3.6)$$
on the invariant manifold $J^{-1}(0)$, where
$$M(q,p,\xi)(z) = L(q,p,\xi)(z)/z. \eqno(2.3.7)$$
\endproclaim
\smallskip
Clearly, the second part of the above proposition is a consequence of
Theorem 2.2.1 (b) and Theorem 2.3.3.

In order to discuss the associated integrable models, we have to
restrict to a smooth component of the reduced Poisson variety
$J^{-1}(0)/H = U \times \fh\times (\fh^{\perp}/H)$.  For this
purpose, we restrict to the following open submanifold of $\fg$:
$${\Cal U} =\{\,\xi\in \fg \mid {\xi}_{\alpha_i} = (\xi, e_{-\alpha_i})
\neq 0, \quad i=1,\ldots, N \,\}. \eqno(2.3.8)$$
Then the $H$-action in Proposition 2.3.5 above induces a Hamiltonian
$H$-action on $TU\times {\Cal U}$ and we denote the corresponding
momentum map also by $J$ so that
$J^{-1}(0) = TU \times (\fh^{\perp}\cap {\Cal U})$.
Now, recall from \c{LX2} that the formula
$$g(\xi ) =\exp{\left (\sum_{i=1}^{N}
\sum_{j=1}^{N}(C_{ji} \log{\xi_{\alpha_{j}}})h_{\alpha_{i}}\right)}\eqno(2.3.9)$$
defines an $H$-equivariant map $g: {\Cal U}\longrightarrow H$ where
$C= (C_{ij})$ is the inverse of the Cartan matrix and 
$h_{\alpha_{i}} = {2 \over (\alpha_i,\alpha_i)} H_{\alpha_{i}}$, 
$i=1,\ldots, N$.  Using $g$, we can identify the reduced space
$J^{-1}(0)/H = TU \times (\fh^{\perp} \cap {\Cal U}/H)$ with
$TU\times \fg_{red}$, where $\fg_{red} = \epsilon + 
\sum_{\alpha \in \Delta - \pi} {\Bbb C} e_{\alpha}$,
and $\epsilon = \sum_{j=1}^{N} e_{\alpha_{j}}$.  Thus the projection
map $\pi_{0} : J^{-1}(0) \longrightarrow TU\times \fg_{red}$
is the map
$$(q,p, \xi)\mapsto  (q,p, Ad_{g(\xi)^{-1}} \xi).\eqno(2.3.10)$$

We shall write $s = \sum_{\alpha\in \Delta} s_{\alpha} e_{\alpha}$
for $s\in \fg_{red}$ (note that
$s_{\alpha_{j}} = 1\,\, \hbox{for}\, j= 1,\ldots, N$).
By Poisson reduction \c{MR}, the reduced manifold $TU\times \fg_{red}$
has a unique Poisson structure which is a product structure where
the second factor $\fg_{red}$ is equipped with the reduction
(at 0) of the Lie-Poisson structure on ${\Cal U}$ by the $H$-action.
If ${\Cal H}$ is the Hamiltonian defined in (2.3.5), we shall
denote its reduction to $TU\times \fg_{red}$ by ${\Cal H}_{0}.$

\bigskip
\bigskip

\subhead
3. \ The rational spin Calogero-Moser systems
\endsubhead

\medskip
\subhead
3.1. \ Lax operators, Hamiltonian equations and the Lie subalgebroids
\endsubhead

\bigskip
Let $\fg$ be a complex simple Lie algebra, as in Section 2.3.  
In addition to the basis $\{e_{\alpha}\}_{\alpha\in \Delta}$
of $\sum_{\alpha\in \Delta} g_{\alpha}$ in that section, let us now
fix an orthonormal basis $(x_i)_{1\le i\le N}$ of $\fh$.  Thus
we will write $p=\sum_{i} p_{i} x_{i}$, \, \,$\xi=\sum_{i} \xi_{i} x_{i} +
\sum_{\alpha \in \Delta} \xi_{\alpha} e_{\alpha}$ for
$p \in \fh$ and $\xi \in \fg.$

The rational spin Calogero-Moser systems are the Hamiltonian systems
on $TU\times \fg$  (as defined in Definition 2.3.4) associated to the 
rational dynamical r-matrices with spectral parameter:

$$r(q,z) = {\Omega \over z} + \sum_{\alpha\in \Delta^{\prime}}
  {1 \over \alpha(q)} e_{\alpha} \otimes e_{-\alpha}, \eqno(3.1.1)$$
where $\Delta^{\prime} \subset \Delta$ is any set of
roots which is closed with respect to addition and multiplication
by $-1,$ and $\Omega\in (S^{2}\fg)^{\fg}$ is the Casimir element
corresponding to the Killing form $(\cdot,\cdot).$  Accordingly, the 
Lax operators are given by
$$L(q,p,\xi)(z) = p + \sum_{\alpha \in \Delta^{\prime}} \frac{\xi_{\alpha}} 
  {\alpha(q)} e_{\alpha} + {\xi\over z} \eqno(3.1.2)$$
and so we have a family of Hamiltonians parametrized by $\Delta^{\prime}$:
$${\Cal H}(q,p,\xi) = {1\over 2} \sum_{i} p_{i}^{2} - {1\over 2}
  \sum_{\alpha \in \Delta^{\prime}} \frac{\xi_{\alpha}\xi_{-\alpha}}
  {\alpha(q)^{2}}.\eqno(3.1.3)$$
Note that in particular, we have
$$L(q,p,\xi)(\infty) \in \fg_{\Delta^{\prime}} \eqno(3.1.4)$$
where
$$\fg_{\Delta^{\prime}} = \fh + \sum_{\alpha\in \Delta^{\prime}} \fg_{\alpha}
  \eqno(3.1.5)$$
is a reductive Lie subalgebra of $\fg$.  As the reader will see, this
fact is important later on, when we solve the factorization problem.

\proclaim
{Proposition 3.1.1} The Hamiltonian equations of motion generated
by ${\Cal H}$ on $TU\times \fg$ are given by
$$\eqalign{& \dot q = p, \cr
           & \dot p = - \sum_{\alpha \in \Delta^{\prime}}
             \frac{\xi_{\alpha} \xi_{-\alpha}} 
              {\alpha(q)^{3}} H_{\alpha},\cr
           & \dot \xi = \left[\,\xi,- \sum_{\alpha \in \Delta^{\prime}}
             \frac{\xi_{\alpha}} {\alpha (q)^{2}}
             e_{\alpha}\,\right].\cr}\eqno(3.1.6)$$
\endproclaim

\demo
{Proof} From the expression
$$\{\varphi,\psi\}(q,p,\xi) =(\delta_{2}\varphi,\delta_{1}\psi)
-(\delta_{1}\varphi, \delta_{2}\psi)+ (\xi,[\delta\varphi,\delta\psi])$$
for the Poisson bracket on $TU\times \fg$, 
the equations of motion are given by
$\dot q = \delta_{2} {\Cal H}$, $\dot p = -\delta_{1} {\Cal H}$,
and $\dot \xi = [\xi, \delta {\Cal H}]$.  Therefore, (3.1.6)
follows by a direct computation.
\pf
\enddemo

We shall solve these equations on $J^{-1}(0)$ by our factorization method.  
To do so, it is essential to have the explicit expression of the classical
dynamical r-matrix $R$ associated to $r$.  Before we make the computation,
let us recall
that the loop algebra $L\fg$ admits a direct sum decomposition
$$L\fg = L^{+}\fg \oplus L^{-}_{0} \fg \eqno(3.1.7)$$
into Lie subalgebras, where $L^{+}\fg$ consists of convergent
power series $\sum_{0}^{\infty} X_{n}z^{n}$, while 
$L^{-}_{0} \fg$ consists of Laurent tails, of the form
$\sum_{-\infty}^{-1} X_{n}z^{n}$.  We shall denote by
$\Pi_{\pm}$ the projection operators relative to this splitting.

\proclaim
{Proposition 3.1.2}  The classical dynamical r-matrix R associated with
the meromorphic map $r$ in (3.1.1) is given by
$$(R(q)X)(z) = {1\over 2}\bigl(\Pi_{+} X - \Pi_{-} X\bigr)(z)
  - \sum_{\alpha \in \Delta^{\prime}} \frac{(X_{-1})_{\alpha}}
  {\alpha(q)} e_{\alpha}. \eqno(3.1.8)$$
In particular,
$$(R(q)M(q,p,\xi))(z) = -{1\over 2}M(q,p,\xi)(z) -\sum_{\alpha\in
  \Delta^{\prime}} \frac{\xi_{\alpha}}{{\alpha(q)}^2} e_{\alpha}\eqno(3.1.9)$$
for $M(q,p,\xi)(z) = L(q,p,\xi)(z)/z.$ (See (2.3.6),(2.3.7).)
\endproclaim
\demo
{Proof} By direct differentiation, we find that 
$ \frac{\partial^{k}r}{\partial z^{k}} (q,-z) = - k! \frac{\Omega}
  {z^{k+1}}$, $k \geq 1$.  Substituting into (2.3.1), the formula
follows.   \pf
\enddemo

\noindent {\bf Remark 3.1.3.} (a) The formula in (3.1.8) shows that the
classical dynamical r-matirx $R$ is a perturbation of the standard
r-matrix associated with the splitting in (3.1.7).
\smallskip
\noindent (b) If we restrict ourselves to $J^{-1}(0)$, then by
equating the coefficients of $z^{0}$ and $z^{-1}$ on both sides
of the Lax equation 
$\dot L (q,p,\xi) = [\, L(q,p,\xi), R(q) M(q,p,\xi) \,]$,
we can recover the equations for $p$ and $\xi$ respectively in (3.1.6).
However, the Lax equation only gives
$\alpha(\dot q - p)$ for all $\alpha \in \Delta^{\prime}$.  Therefore,
unless $\Delta^{\prime} = \Delta$, otherwise, we cannot recover
the equation for $q$ from that of $L(q,p,\xi)$.  This remark shows
that the full set of equations in (2.3.6) is important.
\smallskip
We now give the equations of motion for the reduction of ${\Cal H}$
on $TU \times \fg_{red}$, with Hamiltonian given by
$${\Cal H}_{0}(q,p,s) = {1\over 2} \sum_{i} p_{i}^{2} - {1\over 2}
  \sum_{\alpha \in \Delta^{\prime}} \frac{s_{\alpha} s_{-\alpha}}
  {\alpha(q)^{2}}.\eqno(3.1.10)$$

\proclaim
{Proposition 3.1.4} The Hamiltonian equations of motion generated by
${\Cal H}_0$ on the reduced Poisson manifold $TU\times \fg_{red}$
are given by
$$\aligned
         & \dot q = p, \\
         & \dot p = - \sum_{\alpha \in \Delta^{\prime}}
             \frac{s_{\alpha} s_{-\alpha}} 
              {\alpha(q)^{3}} H_{\alpha},\cr \\
         & \dot s = [\,s, {\Cal M}\,] 
\endaligned
$$
where 
$${\Cal M} =  - \sum_{\alpha \in \Delta^{\prime}}
           \frac{s_{\alpha}} {\alpha (q)^2} e_{\alpha}
          + \sum_{i,j} C_{ji}
           \sum \Sb \alpha \in \Delta^{\prime} \\
           \alpha_{j}-\alpha \in \Delta \endSb N_{\alpha, \alpha_{j}-\alpha} 
           \frac{s_{\alpha} s_{\alpha_{j}-\alpha}} 
           {\alpha(q)^2}  
           h_{\alpha_i}.$$
(Here we use the notation $[e_{\alpha}, e_{\beta}] = N_{\alpha,\beta}\,
e_{\alpha +\beta}$ if $\alpha + \beta \in \Delta$.)
\endproclaim

\demo
{Proof} The equations for $q$ and $p$ are obvious from Propostion 3.1.1
and the fact that $s_{\alpha} = \xi_{\alpha}e^{-\alpha(log g(\xi))}.$
To derive the equation for $s$, we differentiate
$s = Ad_{g(\xi)^{-1}} \xi$ with respect to $t$, assuming that
$\xi$ satisfies the equation in Proposition 3.1.1 with
$\Pi_{\fh} \xi =0.$  This gives
$$\dot s = \left[s, - \sum_{\alpha \in \Delta^{\prime}}
             \frac{s_{\alpha}} { \alpha (q)^{2}}
             e_{\alpha} -T_{g(\xi)^{-1}}r_{g(\xi)} {d\over dt} g(\xi)^{-1}\,\right ].
\qquad (*)$$
Now, from the expression for $g(\xi)$ in (2.3.9), we find
$$-T_{g(\xi)^{-1}}r_{g(\xi)} {d\over dt} g(\xi)^{-1}=\sum_{i,j} C_{ji}\dot\xi_{\alpha_j}
  {\xi_{\alpha_j}}^{-1} h_{\alpha_i}.$$
But from Proposition 3.1.1, we have
$$\aligned
\dot \xi_{\alpha_j}=& \left(\left[\xi, - \sum_{\alpha \in \Delta^{\prime}}
             \frac{\xi_{\alpha}} {\alpha (q)^{2}}e_{\alpha}\right], e_{-\alpha_j}
             \right)\\
=&\,\,\xi_{\alpha_j}\sum \Sb \alpha \in \Delta^{\prime} \\
           \alpha_{j}-\alpha \in \Delta \endSb N_{\alpha, \alpha_{j}-\alpha} 
           \frac{s_{\alpha} s_{\alpha_{j}-\alpha}}{\alpha(q)^2}.\\
\endaligned
$$
Therefore, on substituting the above expressions into (*), we
obtain the desired equation for $s$.
\pf
\enddemo

In order to solve the equations in (2.3.6) by the factorization method, it
is necessary to have precise description of the Lie algebroids and Lie
groupoids which are involved.  We now begin to describe these geometric
objects.  Let $L^{-}\fg$ be the Lie subalgebra of $L\fg$ consisting of
series of the form $\sum_{-\infty}^{0} X_{n} z^{n}$.  From the 
explicit expression for $R$ in (3.1.8), we have
$$(R^{\pm}(q)X)(z) = \pm(\Pi_{\pm}X)(z) - \sum_{\alpha \in \Delta^{\prime}} 
  \frac{(X_{-1})_{\alpha}} {\alpha(q)} e_{\alpha}.\eqno(3.1.11)$$
Therefore, $R^{+}(q)X \in L^{+}\fg$, while 
$R^{-}(q)X \in L^{-}_{\Delta^{\prime}} \fg$, where
$$ L^{-}_{\Delta^{\prime}} \fg = \{ X\in L^-\fg\mid X(\infty) \in
   \fg_{\Delta^{\prime}} \}. \eqno(3.1.12)$$
The proof of the next proposition is obvious and will be
left to the reader.

\proclaim
{Proposition 3.1.5} (a) $Im {\Cal R}^{+} = \bigcup_{q\in U} \{0_q\} \times
L^{+}\fg \times \fh$.
\smallskip
\noindent (b) ${\Cal I}^{+} = \bigcup_{q\in U} \{0_q\} \times L^{+}\fg \times 
\{0\}$ = adjoint bundle of $Im {\Cal R}^+$.
\endproclaim
\smallskip
\noindent{\bf Remark 3.1.6.} Indeed, we also have
$\left \{{\Cal R}^{+}(0_q,X,0)\mid q\in U, X\in L\fg \right\}
 = \bigcup_{q\in U} \{0_q\} \times L^{+}\fg \times \fh.$
\smallskip
Before we turn to the characterization of $Im {\Cal R}^-$, let us recall 
the notion of
a matched pair of Lie algebroids  introduced in \c{Mok} as an
infinitesimal version of the notion of a matched pair of Lie groupoids
\c{M2}.(These are generalizations of the corresponding notions for Lie
algebras and Lie groups, see \c{KSM}, \c{LW},\c{Maj}.)

\definition
{Definition 3.1.7} Two Lie algebroids $A_{1}$, $A_{2}$ over the 
base $B$ is said to form a matched pair of Lie algebroids iff the
Whitney sum $W = A_{1} \oplus A_{2}$ admits a Lie algebroid structure
over the same base with $A_{1}$ and $A_{2}$ as Lie subalgebroids.  In
this case, the Lie algebroid $W$ is called the matched product of
$A_{1}$ and $A_{2}$ and is denoted by $A_{1}\bowtie A_{2}$.
\enddefinition

\proclaim
{Proposition 3.1.8} (a) The ideal ${\Cal I}^{-}$ is given by
$${\Cal I}^{-} = \left\{(0_q,X,0)\Bigm| q\in U, X\in L^{-}_{\Delta^{\prime}}\fg,\,
  X_{-1} \in \fh^{\perp} \,\,\hbox{and}\,\, \sum_{\alpha \in \Delta^{\prime}}
  (X_{-1})_{\alpha} e_{\alpha} = ad_{q} \Pi_{\fh^{\perp}} X_{0}\right \}.$$
\smallskip
\noindent (b) $Im {\Cal R}^{-} = {\Cal I}^{-} \bowtie {\Cal Q}$, 
where
$${\Cal Q} = \bigl \{(0_q,-\widetilde Z, Z)\mid q\in U, Z\in \fh,\,\,
  \hbox {and}\,\,\,\widetilde Z(z) = Z z^{-1} \bigr \}$$
is a Lie subalgebroid of $Im {\Cal R}^-$.
Hence ${\Cal I}^{-}$ coincides with the adjoint bundle of the transitive
Lie algebroid $Im {\Cal R}^{-}$.  Moreover,
${\Cal R}^{-}(\{0_q\}\times L\fg \times \{0\})$ can be characterized
as the set
$$\left \{(0_q,X,Z)\mid Z\in \fh, X\in L^{-}_{\Delta^{\prime}}
          \fg, \Pi_{\fh}X_{0} =0 \,\,\hbox {and}\,\,\, 
          \Pi_{\fg_{\Delta^{\prime}}} X_{-1} = -Z + ad_{q}X_{0} \right \}$$
where $\Pi_{\fg_{\Delta^{\prime}}}$ is the projection map relative to
the decomposition $\fg=\fg_{\Delta^{\prime}}\oplus (\fg_{\Delta^{\prime}})^{\perp}.$
\endproclaim

\demo
{Proof} (a) From the definition of ${\Cal I}^{-}$ in (2.1.10) and 
the expression for $R^{+}(q)$ in (3.1.11), we have
$$\aligned
         & (0_q,X,0) \in {\Cal I}^{-} \\
    \iff & {\Cal R}^{+}(0_q,X,Z) = 0 \,\,\hbox {for some}\,\, Z \in \fh \\
    \iff & \Pi_{\fh} X_{-1} = 0, \,\, -\iota Z + (\Pi_{+}X)(z) - 
            \sum_{\alpha \in \Delta^{\prime}} 
            \frac{(X_{-1})_{\alpha}} {\alpha(q)} e_{\alpha} = 0
           \,\hbox{ for some} \,\, Z\in \fh \\
    \iff & X \in  L^{-}_{\Delta^{\prime}}\fg,\, X_{-1} \in \fh^{\perp}
           \,\,\hbox{and}\,\,  \sum_{\alpha \in \Delta^{\prime}} 
           (X_{-1})_{\alpha}
           e_{\alpha} = ad_{q} \Pi_{\fh^{\perp}} X_{0}.\\
\endaligned $$
Hence the assertion.
\newline
(b) For an element $Z\in \fh$, let  $\widetilde Z$ be the loop
given by  $\widetilde Z(z) = Zz^{-1}$.  Consider an arbitrary element
${\Cal R}^{-}(0_q,X,Z)$ in $Im {\Cal R}^-$.  Clearly, it admits the
decomposition
$$\aligned
         & {\Cal R}^{-}(0_q,X,Z) \\
     =\, & (0_q, -\iota Z - \Pi_{-}X + \widetilde {\Pi_{\fh} X_{-1}}
           - \sum_{\alpha \in \Delta^{\prime}} 
            \frac{(X_{-1})_{\alpha}} {\alpha(q)} e_{\alpha}, 0)\\
         & + (0_q, -\widetilde {\Pi_{\fh} X_{-1}},\Pi_{\fh} X_{-1})\\
\endaligned $$
where the first term is in ${\Cal I}^-$ and the second term is in
${\Cal Q}$.  This shows that 
$Im {\Cal R}^{-} \subset {\Cal I}^{-} \oplus {\Cal Q}.$  Conversely,
take an arbitrary element 
$(0_q, X, 0) + (0_q,-\widetilde Z, Z) \in {\Cal I}^{-} \oplus {\Cal Q}$
and let $Y\in L\fg$ be defined by $Y = -\Pi_{-}X + \widetilde Z.$
Then from the characterization of ${\Cal I}^{-}$ in part (a), we
have
$$\aligned
         &{\Cal R}^{-}(0_q,Y, -\Pi_{\fh} X_{0}) \\
     =\, &(0_q,\Pi_{\fh} X_{0} -\Pi_{-}Y - \sum_{\alpha \in \Delta^{\prime}} 
            \frac{(Y_{-1})_{\alpha}} {\alpha(q)} e_{\alpha},
            \Pi_{\fh} Y_{-1})\\
     =\, & (0_q,\Pi_{\fh} X_{0} +\Pi_{-}X-\widetilde Z + 
            \sum_{\alpha \in \Delta^{\prime}} 
           (X_{0})_{\alpha}  e_{\alpha}, Z)\\
     =\, & (0_q, X, 0) + (0_q,-\widetilde Z, Z) \\
\endaligned $$
and this shows 
${\Cal I}^{-} \oplus {\Cal Q}\subset Im {\Cal R}^{-}.$  Combining the
two inclusions, we conclude that 
$Im {\Cal R}^{-} = {\Cal I}^{-} \bowtie {\Cal Q}.$ We shall
leave the details of the other assertions to the reader.
\pf
\enddemo

As a consequence of Proposition 3.1.5, we have
$$\eqalign{Im {\Cal R}^{+}/{\Cal I}^{+}& = \bigcup_{q\in U} 
  \lbrace 0_q \rbrace \times (L^{+}\fg/L^{+}\fg)\times \fh\cr
  & \simeq \bigcup_{q\in U} \lbrace 0_q \rbrace \times
    \lbrace 0 \rbrace \times \fh\cr}\eqno(3.1.13)
$$
where the identification map is given by
$$(0_q, X+ L^{+}\fg, Z) \mapsto (0_q,0,Z).\eqno(3.1.14)$$
Similarly, it follows from Proposition 3.1.8 that
$$Im {\Cal R}^{-}/{\Cal I}^{-} \simeq {\Cal Q} \eqno(3.1.15)$$
and the identification map is
$$(0_q,X,Z) + {\Cal I}^{-}_q \mapsto (0_q, -\widetilde Z, Z).\eqno(3.1.16)$$
The following proposition is obvious.

\proclaim
{Proposition 3.1.9} The isomorphism $\theta : Im {\Cal R}^{+}/{\Cal I}^{+}
\longrightarrow Im {\Cal R}^{-}/{\Cal I}^{-}$ defined in Proposition 2.1.2 (b)
is given by
$$\theta (0_q,0,Z) = (0_q,-\widetilde Z, Z).$$
Moreover, 
$Im ({\Cal R}^{+}, {\Cal R}^{-}) =Im{\Cal R}^{+}
{\underset TU \to\oplus}Im{\Cal R}^{-}$.
\endproclaim
\medskip

\subhead
3.2. Solution of the integrable rational spin Calogero-Moser systems
\endsubhead

\medskip
We begin by solving the equation
$$\eqalign{
  &{d\over dt} \, (q, 0, L(q,p,\xi))\cr
 =\,&(p, 0, [\,L(q,p,\xi), R(q) M(q,p,\xi)\,])\cr}\eqno(3.2.1)
$$
where explicitly,
$$M(q,p,\xi)(z) = {1\over z}\left( p + \sum_{\alpha \in \Delta^{\prime}} 
\frac{\xi_{\alpha}} {\alpha(q)} e_{\alpha} \right)+ {\xi\over z^{2}}.\eqno(3.2.2)$$
To do so, we have to solve the factorization problem
$$exp \{\, t(0,0,M(q^0,p^0,\xi^0))\}(q^0)
      =\,\gamma_{+} (t)\,\gamma_{-} (t)^{-1}\eqno(3.2.3)$$
for 
$(\gamma_{+}(t),\gamma_{-}(t))=
((q^0, k_{+} (t), q(t)), (q^0, k_{-} (t), q(t))) \in
Im ({\Cal R}^{+}, {\Cal R}^{-})$
satisfying
the condition in (2.2.4), where
$(q^0,p^0,\xi^0) \in J^{-1}(0) = TU \times ({\Cal U} \cap \fh^{\perp})$
is the initial value of $(q,p,\xi)$.  In what follows, we shall denote
by $LG$, $L^{+}G$, $L^{-}_{1}G$ and $L^{-}_{\Delta^{\prime}}G$ the loop
groups corresponding to the Lie algebras
$L\fg$, $L^{+}\fg$, $L^{-}_{0}\fg$ and $L^{-}_{\Delta^{\prime}}\fg$
respectively.  We shall also denote $k_{\pm}(t)(z)$ by
$k_{\pm}(z,t)$.  Then $k_{+}(\cdot,t)\in L^{+}G$, while
$k_{-}(\cdot,t)\in L^{-}_{\Delta^{\prime}}G$ and satisfies 
additional constraints.  From the factorization problem on the
Lie groupoid above, it follows that
$$e^{t M(q^0,p^0,\xi^0)(z)} = k_{+}(z,t)k_{-}(z,t)^{-1} \eqno(3.2.4)$$
where $k_{\pm}(\cdot,t)$ are to be determined.  To do so, we recall
from the Birkhoff factorization theorem \c{PS} that (at least for
small values of $t$)
$$e^{t M(q^0,p^0,\xi^0)(z)} = g_{+}(z,t)g_{-}(z,t)^{-1} \eqno(3.2.5)$$
for unique $g_{+}(\cdot,t)\in L^{+}G$ and $g_{-}(\cdot,t)\in L^{-}_{1}G$.
But from (3.2.2) above, it is clear that $e^{t M(q^0,p^0,\xi^0)} \in L^{-}_{1}G$, 
so we must have
(i) $g_{+} \equiv 1$, (ii) $g_{-}(z,t) = e^{-t M(q^0,p^0,\xi^0)}(z)$ for
all $t$ and consequently,
$$k_{+}(z,t)\equiv k_{-} (\infty,t). \eqno(3.2.6)$$
Thus we have the relation 
$$e^{t M(q^0,p^0,\xi^0)(z)} = k_{-}(\infty,t) k_{-}(z,t)^{-1} \eqno(3.2.7)$$
where $\gamma_{-}(t) = (q^0,k_{-}(t), q(t))$ is subject to the condition
$T_{\gamma_{-} (t)} {\bbl}_{{\gamma_{-}(t)}^{-1}} {\dot\gamma_{-}(t)}
\in {\Cal R}^{-}(\lbrace q(t) \rbrace \times \lbrace 0 \rbrace \times L\fg).$
But from the characterization of 
${\Cal R}^{-}(\lbrace q(t) \rbrace \times \lbrace 0 \rbrace \times L\fg)$ 
in Proposition 3.1.8 (b), we have
$$\eqalign{ 
         & \Pi_{\fg_{\Delta^{\prime}}}\, Res_{z=0}\, T_{k_{-}(z,t)}
           l_{k_{-}(z,t)^{-1}}\dot k_{-}(z,t)\cr
     =\, & -\dot q(t) + ad_{q(t)}T_{k_{-}(\infty,t)}l_{k_{-}(\infty,t)^{-1}}
           \dot k_{-}(\infty,t).\cr}\eqno(3.2.8)$$
On the other hand, by differentiating (3.2.7) with respect to $t$, we
find
$$
\aligned
       & T_{k_{-}(z,t)}l_{{k_{-}(z,t)}^{-1}} {\dot k_{-}(z,t)}
        - T_{k_{-}(\infty,t)}l_{{k_{-}(\infty,t)}^{-1}}{\dot k_{-}(\infty,t)}\\
    =\,& -Ad_{k_{-}(\infty,t)^{-1}} M(q^0,p^0,\xi^0)(z)
\endaligned
$$
from which it follows that
$$Res_{z=0}\, T_{k_{-}(z,t)}l_{{k_{-}(z,t)}^{-1}} {\dot k_{-}(z,t)}
 = -Ad_{k_{-}(\infty,t)^{-1}} \left (p^0 + \sum_{\alpha\in \Delta^{\prime}}
   \frac{\xi^{0}_{\alpha}}{\alpha(q^0)} e_{\alpha} \right ).\eqno(3.2.9)$$
Therefore, upon substituting (3.2.9) into (3.2.8), we obtain
$$\aligned
         & Ad_{k_{-}(\infty,t)}\, \dot q(t) + \left [\,T_{k_{-}(\infty,t)}
           r_{{k_{-}(\infty,t)}^{-1}}\dot k_{-}(\infty,t),
           Ad_{k_{-}(\infty,t)}\, q(t)\, \right ] \\
     =\, & L(q^0,p^0,\xi^0)(\infty), \\
\endaligned
$$
that is,
$${d\over dt} Ad_{k_{-}(\infty,t)}\, q(t) =  L(q^0,p^0,\xi^0)(\infty).\eqno(3.2.10)$$
Hence the factorization problem boils down to
$$q^0 + t L(q^0,p^0,\xi^0)(\infty) = Ad_{k_{-}(\infty,t)}\, q(t)\eqno(3.2.11)$$
where $q(t)$ and $k_{-}(\infty,t)$ are to be determined.  But from
(3.1.4) and the fact that $\fg_{\Delta^{\prime}}$ is reductive, we can
find (at least for small values of $t$) unique $d(t)\in H$ and
$g(t)\in G_{\Delta^{\prime}}$ (unique up to $g(t) \to g(t)\delta (t)$,
where $\delta (t) \in H$) such that
$$q^0 + t L(q^0,p^0,\xi^0) = Ad_{g(t)}\, d(t)\eqno(3.2.12)$$
with $g(0) =1$, $d(0) =q^0$.  Hence
$$q(t) = d(t). \eqno(3.2.13)$$
On the other hand, let us fix one such $g(t)$.  We shall seek
$k_{-}(\infty,t)$ in the form
$$k_{-}(\infty,t) = g(t)h(t), \quad h(t)\in H. \eqno(3.2.14)$$
To determine $h(t)$, note that the characterization of
${\Cal R}^{-}(\lbrace q(t) \rbrace \times \lbrace 0 \rbrace \times L\fg)$
in Proposition 3.1.8 (b) also gives
$$\Pi_{\fh}\, T_{k_{-}(\infty,t)}l_{{k_{-}(\infty,t)}^{-1}}\dot k_{-}(\infty,t)
  = 0. \eqno(3.2.15)$$
Using this condition, we find that $h(t)$ satisfies the equation
$$\dot h(t) = T_{e} l_{h(t)}\left ( - \Pi_{\fh} 
(T_{g(t)} l_{{g(t)}^{-1}} \dot g(t))\right )\eqno(3.2.16)$$
with $h(0) =1$.  Solving the equation explicitly, we obtain
$$h(t) =  exp \left\{-\int _{0}^{t}\Pi_{\fh} 
(T_{g(\tau)} l_{{g(\tau)}^{-1}} 
\dot g(\tau))\, d\tau \right\} .\eqno(3.2.17)$$
Hence $k_{+}(z,t)\equiv k_{-}(\infty,t)$ and 
$k_{-}(z,t)\equiv e^{-t M(q^0,p^0,\xi^0)(z)} k_{-}(\infty,t)$
satisfy (3.2.4).

\proclaim
{Theorem 3.2.1} Let 
$(q^0,p^0,\xi^0) \in J^{-1}(0) = TU \times ({\Cal U} \cap \fh^{\perp}).$
Then the Hamiltonian flow on $J^{-1} (0)$ generated by
$${\Cal H}(q,p,\xi) = {1\over 2} \sum_{i} p_{i}^{2} - {1\over 2}
  \sum_{\alpha \in \Delta^{\prime}} \frac{\xi_{\alpha}\xi_{-\alpha}}
  {\alpha(q)^{2}}$$

with initial condition $(q(0), p(0), \xi(0)) = (q^0,p^0,\xi^0)$ is
given by
$$\eqalign{& q(t) =\, d(t), \cr
       & \xi (t) =\, Ad_{k_{-}(\infty,t)^{-1}} \xi^{0},\cr
       & p(t) =\,\dot d(t) =\, Ad_{k_{-}(\infty,t)^{-1}} L(q^0,p^0,\xi^0)
         (\infty)- \sum_{\alpha \in \Delta^{\prime}}
         \frac{\xi(t)_{\alpha}}{\alpha (q(t))}  e_{\alpha}\cr}\eqno(3.2.18)
$$
where $d(t)$ and $k_{-}(\infty,t)$ are constructed from the above
procedure.
\endproclaim

\demo
{Proof} The formulas for $p(t)$, $\xi(t)$ are obtained by equating
the coefficients of $z^0$ and $z^{-1}$ on both sides of the
expression
$L(q(t),p(t),\xi(t))(z) = Ad_{k_{-}(\infty,t)^{-1}} L(q^0,p^0,\xi^0)(z).$
\pf
\enddemo

We now turn to the solution of the associated integrable model on
$TU\times \fg_{red}$ with Hamiltonian
${\Cal H}_{0} (q,p,s) = {1\over 2} \sum_{i} p^{2} -{1\over 2}
 \sum_{\alpha\in \Delta^{\prime}}\frac{ s_{\alpha} s_{-\alpha}}
 {\alpha(q)^{2}}$ and with equations of motion given in Proposition 3.1.4.

\proclaim
{Corollary 3.2.2} Let $(q^0,p^0,s^0) \in TU \times \fg_{red}$ and
suppose $s^0 =  Ad_{g(\xi^{0})^{-1}} \xi^{0}$ where 
$\xi^{0}\in {\Cal U} \cap \fh^{\perp}.$  Then the Hamiltonian
flow generated by ${\Cal H}_{0}$ with initial condition
$(q(0),p(0),s(0)) = (q^0,p^0,s^0)$ is given by
$$\eqalign{
       & q(t) =\, d(t), \cr
       & s(t) =\, Ad_{\bigl({\widetilde k}_{-}(\infty,t)\,
                  g \bigl(Ad_{{\widetilde k}_{-}(\infty,t)^{-1}}\, s^{o}\bigr)
       \bigr)^{-1}} s^{0},\cr
       & p(t) =\, Ad_{\bigl({\widetilde k}_{-}\,(\infty,t)\,
                  g\bigl(Ad_{{\widetilde k}_{-}(\infty,t)^{-1}} s^{o}\bigr)
         \bigr)^{-1}} L(q^0,p^0,s^0)(\infty) -\sum_{\alpha \in \Delta^{\prime}}
         \frac{s_{\alpha}(t)}{\alpha (q(t))} e_{\alpha}.\cr}
\eqno(3.2.19)
$$
where ${\widetilde k}_{-}(\infty,t) = g(\xi^{0})^{-1} k_{-}(\infty,t) g(\xi^{0})$
depends only on $s^0$ and $k_{-}(\infty,t)$, $d(t)$ are as in Theorem 3.2.1.
\endproclaim

\demo
{proof} We shall obtain the Hamiltonian flow generated by ${\Cal H}_{0}$
by Poisson reduction.  Using the relation
$\phi^{red}_{t}\circ\pi_{0} =\pi_{0}\circ\phi_{t}\circ i_{0}$ from
Corollary 2.2.3, we have
$$\phi^{red}_{t}(q^0,p^0,s^0)=(q(t),p(t),Ad_{g(\xi(t))^{-1}}\xi(t))$$
where $q(t)$ and $p(t)$ are given by the expressions in Theorem
3.2.1 above.  Thus 
$$s(t) = Ad_{g(\xi(t))^{-1}}\xi(t)=Ad_{\bigl({\widetilde k}_{-}(\infty,t)\,
                  g \bigl(Ad_{{\widetilde k}_{-}(\infty,t)^{-1}}\, s^{o}\bigr)
             \bigr)^{-1}} s^{0}$$
where we have used the $H$-equivariance of the map $g$ to show that
$$g(\xi^0)^{-1}k_{-}(\infty,t)g(\xi(t)) = {\widetilde k}_{-}\,(\infty,t)\,
                  g\bigl(Ad_{{\widetilde k}_{-}(\infty,t)^{-1}} s^{o}\bigr).$$
To express $p(t)$ in the desired form, simply apply $Ad_{g(\xi(t))^{-1}}$ to
both sides of the expression for $p(t)$ in the above theorem, this gives
$$p(t)=\,Ad_{g(\xi(t))^{-1}k_{-}(\infty,t)^{-1}g(\xi^{0})} L(q^0,p^0,s^0)(\infty)
  -\sum_{\alpha \in \Delta^{\prime}}\frac{s_{\alpha}(t)}{\alpha (q(t))}e_{\alpha}$$
where we have used the relation $s_{\alpha}(t) = e^{-\alpha(g(\xi(t)))}\xi_{\alpha}(t)$.
Hence the desired expression for $p(t)$ follows.  The assertion that
${\widetilde k}_{-}(\infty,t)$ depends only on $s^0$ is clear.
\pf
\enddemo

\noindent{\bf Remark 3.2.3.} (a) The expression in (3.2.12) shows that
the solution blows up precisely when the factorization fails.
However,  initial conditions do exist for which the solution
exists for all time. 
\newline
(b) The first example of a rational spin 
Calogero-Moser system is due to Gibbons and Hermsen \c{GH}.
Analogous to what was done there, we can show that 
$$\dot q = L(q,p,\xi)(\infty) + \left[q\,,-\sum_{\alpha \in \Delta^{\prime}} 
   \frac{\xi_{\alpha}}{\alpha(q)^2} e_{\alpha} \right]$$
on $J^{-1}(0)$
from which we can also deduce the relation (3.2.12).  Thus on the 
surface, it appears that there is no need to use $L(q,p,\xi)(z)$.  
We remark, however,
that our factorization problem (which involves $L(q^0,p^0,\xi^0)(z)$)
does carry more information and
that we do need $L(q,p,\xi)(z)$ in order to establish the Liouville
integrability of the reduced system on $TU\times \fg_{red}.$
In other words, our realization picture embraces both exact
solvability and complete integrability. 
A unifying and representation independent method to establish
the Liouville integrability of the integrable spin CM systems
in \c{LX2} for all simple Lie algebras will be given in a forthcoming paper.
\newline
(c) We now explain the Poisson meaning of the limiting Lax operator 
$L(q,p,\xi)(\infty).$  To do so, we recall that 
$r(q) = \sum_{\Delta^{\prime}} {1\over \alpha(q)} e_{\alpha}\otimes
 e_{-\alpha}$
is a classical dynamical r-matrix with zero coupling constant in
the sense of \c{EV}.  Therefore, if we define 
$R: U\longrightarrow L(\fg,\fg)$ by
$$R(q)\xi = r^{\sharp}(q)\xi =-\sum_{\alpha^{\prime}}\frac{\xi_{\alpha}}
  {\alpha(q)} e_{\alpha},$$
then $R$ is a solution of the CDYBE (i.e., (2.1.2) with $\chi\equiv 0$).
Let $A^{*}\Omega \simeq TU\times \fg$ be the coboundary dynamical
Lie algebroid associated with $R$ and let $A\Omega\simeq TU\times \fg$ 
be the trivial Lie algebroid.  Then according to \c{L2},
$${\Cal R}: A^*\Omega \longrightarrow A\Omega, 
(q, p, \xi)\mapsto (q, \Pi_{\fh}\xi, -p + R(q)\xi)$$
is a morphism of Lie algebroids.  Consequently, the dual map
${\Cal R}^*$ is an $H$-equivariant Poisson map, when the
domain and target are equipped with the corresponding Lie-Poisson
structure.  Explicitly,
$$\aligned
{\Cal R}^{*}(q,p,\xi) =& (q, -\Pi_{\fh}\xi, p - R(q)\xi)\\
                      =& (q, -\Pi_{\fh}\xi, L(q,p,\xi)(\infty)).\\
\endaligned
$$ 
Moreover, if we define $$L^{\infty}(q,p,\xi) =  L(q,p,\xi)(\infty),$$
then the Hamiltonian ${\Cal H}$ of the rational spin CM system
in (3.1.3) is also given by
$${\Cal H}(q,p,\xi) = ((L^{\infty})^{*}E) (q,p,\xi)$$
where $E$ is the quadratic function on $\fg$ defined by
$$E(\xi) = {1\over 2}(\xi,\xi).$$
This shows that the Hamiltonian system defined by ${\Cal H}$
admits a second realization in $A\Omega$ and this clarifies
the Poisson-geometric meaning of $L(q,p,\xi)(\infty)$.  We
note, however, that the r-matrix ${\Cal R}$ introduced
earlier in this remark is degenerate in the sense that it is not associated
with a factorization problem. 

\bigskip
\bigskip

\subhead
4. \ The trigonometric spin Calogero-Moser systems
\endsubhead
\medskip
\subhead
4.1. \ Lax operators, Hamiltonian equations and the Lie subalgebroids
\endsubhead
\medskip

In this section, we take the trigonometric spin Calogero-Moser systems 
to be the Hamiltonian systems in Definition 2.3.4 associated to the 
following trigonometric
dynamical r-matrices with spectral parameter:
$$r(q,z) = c(z)\sum_{i} x_{i}\otimes x_{i} -\sum_{\alpha\in \Delta}
  \phi_{\alpha}(q,z) e_{\alpha}\otimes e_{-\alpha} \eqno(4.1.1)$$
where 
$$c(z) = \cot z  \eqno(4.1.2)$$
and
$$
\phi_{\alpha}(q,z)=\cases -\frac{\sin(\alpha(q)+z)}{\sin\alpha(q)\sin z}, 
  & \alpha\in <\pi^{\prime}> \\
 -\frac{e^{-iz}}{\sin z} , & \alpha\in
  {\overline \pi^{\prime}}^+ \\
 -\frac{e^{iz}}{\sin z}, & \alpha\in
  {\overline \pi^{\prime}}^- .\endcases \eqno(4.1.3)
$$
In (4.1.3) above, $\pi^{\prime}$ is an arbitrary subset of
a fixed simple system $\pi\subset \Delta$, $<\pi^{\prime}>$ is the root
span of $\pi^{\prime}$ and ${\overline
\pi^{\prime}}^{\pm}= \Delta^\pm \setminus <\pi^{\prime}>^\pm.$
Accordingly, the Lax operators are given by
$$\eqalign{
L(q,p,\xi)(z) =\, &p+ c(z)\sum_{i}\xi_{i}x_{i} -
  \sum_{\alpha\in \Delta} \phi_{\alpha}(q,z)\xi_{\alpha}e_{\alpha}\cr
   =\,& p + \sum_{\alpha\in <\pi^{\prime}>} c(\alpha(q))\xi_{\alpha}
      -i\sum_{\alpha\in {\overline\pi^{\prime}}^+} \xi_{\alpha} e_{\alpha}\cr
    &+i\sum_{\alpha\in {\overline\pi^{\prime}}^-} \xi_{\alpha} e_{\alpha} + c(z)\xi.\cr}
    \eqno(4.1.4)$$
Hence we have a family of dynamical systems parametrized by subsets 
$\pi^{\prime}$ of $\pi$ with Hamiltonians of the form:
$$\eqalign{{\Cal H}(q,p,\xi) = &{1\over 2} \sum_{i} p_{i}^{2} - {1\over 2}
  \sum_{\alpha \in <\pi^{\prime}>} \left(\frac{1}{\sin^{2}\alpha(q)}
  -{1\over 3}\right) {\xi_{\alpha}\xi_{-\alpha}}
  -{5\over 6} \sum_{\alpha\in\Delta\setminus <\pi^{\prime}>}
  {\xi_{\alpha}\xi_{-\alpha}}\cr
   & -{1\over 3} \sum_{i} \xi^{2}_{i}.\cr}\eqno(4.1.5)$$

\noindent{\bf Remark 4.1.1.} The trigonometric dynamical r-matrices in
(4.1.1) are gauge equivalent to those used in \c{LX2}. Although the
corresponding Hamiltonians in (4.1.5) above contain the additional term
$-{1\over 3} \sum_{i} \xi^{2}_{i}$, however, the Hamiltonian flows
on $J^{-1}(0)$ and the reduced systems are the same as those in
\c{LX2}.  The reason why we use the dynamical r-matrices in
(4.1.1) is due to the fact that the corresponding Lie subalgebroids
$Im {\Cal R}^{\pm}$ are simpler to analyze.
\smallskip 

The next two propositions follow from direct calculation, as in 
Propositions 3.1.1  and 3.1.4.  We shall leave the proof to the
reader.

\proclaim
{Proposition 4.1.2} The Hamiltonian equations of motion generated
by ${\Cal H}$ on $TU\times \fg$ are given by
$$\eqalign{& \dot q = p, \cr
           & \dot p = - \sum_{\alpha \in <\pi^{\prime}>}
             \frac{\cot \alpha(q)} 
              {\sin^{2}\alpha(q)}\xi_{\alpha} \xi_{-\alpha} H_{\alpha},\cr
           & \dot \xi = \left[\,\xi,-{2\over 3} \Pi_{\fh} \xi-
              \sum_{\alpha \in <\pi^{\prime}>}
             \left (\frac{1} {\sin^{2}\alpha (q)}-{1\over 3}\right)\xi_{\alpha}
             e_{\alpha}-{5\over 3} \sum_{\alpha\in\Delta\setminus 
             <\pi^{\prime}>} \xi_{\alpha} e_{\alpha}\,\right ].\cr}\eqno(4.1.6)$$
\endproclaim

\proclaim
{Proposition 4.1.3}  The Hamiltonian equations of motion generated by
$${\Cal H}_{0}(q,p,s) ={1\over 2} \sum_{i} p_{i}^{2} - {1\over 2}
  \sum_{\alpha \in <\pi^{\prime}>} \left(\frac{1}{\sin^{2}\alpha(q)}
  -{1\over 3}\right) {s_{\alpha}s_{-\alpha}}
  -{5\over 6} \sum_{\alpha\in\Delta\setminus <\pi^{\prime}>}
  {s_{\alpha}s_{-\alpha}}$$
on the reduced Poisson manifold $TU\times \fg_{red}$
are given by
$$\aligned
         & \dot q = p, \\
         & \dot p = - \sum_{\alpha \in \Delta^{\prime}}
             \frac{\cot \alpha(q)} 
              {\sin^{2}\alpha(q)}s_{\alpha}s_{-\alpha} H_{\alpha},\cr \\
         & \dot s = [\,s, {\Cal M}\,] 
\endaligned
$$
where 
$$\aligned
{\Cal M} = & - \sum_{\alpha \in <\pi^{\prime}>}
           \left (\frac{1} {\sin^{2}\alpha (q)}-{1\over 3}\right)s_{\alpha}
             e_{\alpha}-{5\over 3} \sum_{\alpha\in\Delta\setminus 
             <\pi^{\prime}>} s_{\alpha} e_{\alpha}\\
           &  + \sum_{i,j} C_{ji}
           \sum \Sb \alpha \in <\pi^{\prime}>-\pi^{\prime} \\
           \alpha_{j}-\alpha \in \Delta \endSb N_{\alpha, \alpha_{j}-\alpha} 
            \left (\frac{1} {\sin^{2}\alpha (q)}-{1\over 3}\right)s_{\alpha}
            s_{\alpha_{j}-\alpha} h_{\alpha_i}\\
           &+ {5\over 3}\sum_{i,j} C_{ji}
           \sum \Sb \alpha \in\Delta\setminus <\pi^{\prime}> \\
           \alpha_{j}-\alpha \in \Delta \endSb N_{\alpha, \alpha_{j}-\alpha} 
            s_{\alpha}s_{\alpha_{j}-\alpha} h_{\alpha_i}.\\
\endaligned
$$
(Here the notation $N_{\alpha,\beta}$ is as in Proposition 3.1.4.)
\endproclaim

\proclaim
{Proposition 4.1.4} The classical dynamical r-matrix $R$ associated with
the trigonometric dynamical r-matrix with spectral parameter in (4.1.1)
is given by
$$\eqalign{(R(q)X)(z) & = {1\over 2}X(z) +\sum_{k=0}^{\infty} \frac{
           c^{(k)}(-z)}{k!}\,  X_{-(k+1)} -\sum_{\alpha \in <\pi^{\prime}>} 
           c(\alpha(q))(X_{-1})_{\alpha} e_{\alpha} \cr
           & + i\sum_{\alpha \in{\overline \pi^{\prime}}^+}
           (X_{-1})_{\alpha}e_{\alpha}
           -i\sum_{\alpha \in{\overline \pi^{\prime}}^-}(X_{-1})_{\alpha}
            e_{\alpha} .\cr}\eqno(4.1.7)$$
\endproclaim

\demo
{Proof} The formula follows from (2.3.1) and (4.1.1) by a direct calculation 
where we have used the formula 
${d^{k}\over dw^{k}}{\Big|_{w=0}} \phi_{\alpha}(q,z-w)=c^{(k)}(-z),
\,\, k\geq 1$.
\pf
\enddemo

\proclaim
{Corollary 4.1.5} On $J^{-1}(0)$, we have
$$\eqalign{& (R(q)M(q,p,\xi))(z) \cr
       = \,& {1\over 2} M(q,p,\xi)(z) -c(z)p + \sum_{\alpha\in\Delta
            \setminus <\pi^{\prime}>} \phi_{\alpha}(q,z) c(z) \xi_{\alpha}
             e_{\alpha}\cr 
           & +\sum_{\alpha\in <\pi^{\prime}>}
             \phi_{\alpha}(q,z)(c(\alpha(q))+c(z)-c(\alpha(q)+z))\xi_{\alpha}
             e_{\alpha}\cr}\eqno(4.1.8)$$
where $M(q,p,\xi)(z) = L(q,p,\xi)(z)/z.$
\endproclaim

\demo
{Proof} The formula in (4.1.8) follows from (4.1.7) by algebra on
using the following expansion in a deleted neighborhood of 0:
$$ M(q,p,\xi)(z) =\, \frac{\xi}{z^{2}} + {1\over z}\,M(q,p,\xi)_{-1}
   +O(1),$$
where
$$
\aligned
       & M(q,p,\xi)_{-1} \\
    =\,&  p + 
          \sum_{\alpha\in<\pi^{\prime}>} c(\alpha(q))\xi_{\alpha} e_{\alpha}
            - i \sum_{\alpha\in {\overline \pi^{\prime}}^+}
              \xi_{\alpha}e_{\alpha}
         + i \sum_{\alpha\in {\overline \pi^{\prime}}^-}
           \xi_{\alpha}e_{\alpha}.\\
\endaligned
$$
\pf
\enddemo

Our next lemma is obvious from (4.1.7) and the expansion
$c^{(k)}(-z) = -k!z^{-(k+1)} + O(1), k\geq 0,$ in a deleted neighborhood of $0.$

\proclaim
{Lemma 4.1.6} For $X\in L\fg$, $R^{+}(q)X\in L\fg.$
\endproclaim

\proclaim
{Lemma 4.1.7} (a) For $X\in L\fg$, $R^{-}(q)X$
has singularities at the points of the rank one lattice $\pi \Bbb Z$,
and is holomorphic in $\Bbb C \setminus \pi \Bbb Z.$  
Moreover, $R^{-}(q)X$ is simply-periodic with period $\pi$. 
\smallskip
\noindent (b) The principal part of $R^{-}(q)X$ at $z=0$ is 
$-(\Pi_{-}X)(z).$
\smallskip
\noindent (c) $R^{-}(q)X$ is bounded as $z\to\infty$ in a period strip
with 
$$\lim_{y\to\infty}(R^{-}(q) X)(x + iy) = i \Pi_{\fh} X_{-1}
   +\sum_{\alpha\in <\pi^{\prime}>} (i-c(\alpha(q)))(X_{-1})_{\alpha}
   e_{\alpha} +2i \sum_{\alpha \in {\overline \pi^{\prime}}^+} 
    (X_{-1})_{\alpha} e_{\alpha},$$
$$\lim_{y\to\infty}(R^{-}(q) X)(x - iy) = -i \Pi_{\fh} X_{-1}
   -\sum_{\alpha\in <\pi^{\prime}>} (i+c(\alpha(q)))(X_{-1})_{\alpha}
   e_{\alpha} -2i \sum_{\alpha \in {\overline \pi^{\prime}}^-} 
    (X_{-1})_{\alpha} e_{\alpha}.$$
\endproclaim

\demo
{Proof} (a) Clearly, $c^{(k)}(-z)$ are periodic with period $\pi$ and
 meromorphic in $\Bbb C$ with poles at the points of the rank one lattice 
$\pi\Bbb Z$.
Therefore, the assertion follows.
\newline
(b) This follows from the property that for $k\geq 0$, we have 
$c^{(k)}(-z) = -k!z^{-(k+1)} + O(1)$ in a deleted neighborhood
of $z=0$.
\newline
(c) First of all, note that
$\lim_{y\to \pm\infty} \cot(x+iy) = \mp i.$
On the other hand, it is easy to check that the derivatives of
$\cot z$ always contain $\csc^{2} z$ as a factor.  Therefore,
we have 
$\lim_{y\to \pm\infty} c^{(k)}(x+iy) = 0$ for $k\geq 1.$
The formulas for $\lim_{y\to\infty}(R^{-}(q) X)(x \pm iy)$
are now obvious from (4.1.7).
\pf
\enddemo

In order to describe the membership of the elements 
$(R^{-}(q)X)(\pm i \infty)$ in Lemma 4.1.7 (c) and for
subsequent analysis, we need to introduce a number of Lie subalgebras
of $\fg$ and their corresponding Lie groups.  To
begin with, let 
${\frak b}^{-} =\fh + \sum_{\alpha \in \Delta^{-}} \fg_{\alpha}$
and ${\frak b}^{+} =\fh + \sum_{\alpha \in \Delta^{+}} \fg_{\alpha}$
be opposing Borel subalgebras of $\fg.$  Then for each 
$\pi^{\prime}\subset \pi$, we have the parabolic subalgebras
$${\frak p}^{\pm}_{\pi^{\prime}} = {\frak b}^{\pm} + \sum_{\alpha\in
<\pi^{\prime}>^{\mp}}\, \fg_{\alpha}. \eqno(4.1.9)$$
Recall that  ${\frak p}^{\pm}_{\pi^{\prime}}$ admit the following
direct sum decomposition \c{Kn}
$${\frak p}^{\pm}_{\pi^{\prime}} = \fg_{\pi^{\prime}} + 
{\frak n}^{\pm}_{\pi^{\prime}}\eqno(4.1.10)$$
where 
$$\fg_{\pi^{\prime}} = \fh + \sum_{\alpha \in <\pi^{\prime}>} {\fg_{\alpha}}
\eqno(4.1.11)$$
is the Levi factor of ${\frak p}^{\pm}_{\pi^{\prime}}$, and
$${\frak n}^{\pm}_{\pi^{\prime}} = \sum_{\alpha \in
{\overline \pi^{{\prime}^{\pm}}}} \fg_{\alpha} \eqno(4.1.12)$$
are the nilpotent radicals.  We shall denote by 
$\Pi^{\pm}_{\fg_{\pi^{\prime}}}$  the projection maps onto
$\fg_{\pi^{\prime}}$ relative to the splitting
${\frak p}^{\pm}_{\pi^{\prime}} = \fg_{\pi^{\prime}} + 
{\frak n}^{\pm}_{\pi^{\prime}}$.  On the other hand, the connected
and simply-connected Lie subgroups of $G$ with corresponding
Lie subalgebras $\frak{p}^{\pm}_{\pi^{\prime}}$, $\fg_{\pi^{\prime}}$, 
and $\frak{n}^{\pm}_{\pi^{\prime}}$ will be denoted respectively
by $P^{\pm}_{\pi^{\prime}}$, $G_{\pi^{\prime}}$, and $N^{\pm}_{\pi^{\prime}}$
and we have $P^{\pm}_{\pi^{\prime}} =N^{\pm}_{\pi^{\prime}} G_{\pi^{\prime}}$.

Thus it follows from Lemma 4.1.7 (c)
that $(R^{-}(q)X)(\pm i  \infty) \in {\frak p}^{\pm}_{\pi^{\prime}}$. 
The proof of the our next proposition is obvious.
\proclaim
{Proposition 4.1.8}(a) $Im {\Cal R}^{+} = \bigcup_{q\in U} \{0_q\} \times
L^{+}\fg \times \fh$.
\smallskip
\noindent (b) ${\Cal I}^{+} = \bigcup_{q\in U} \{0_q\} \times L^{+}\fg \times 
\{0\}$ = adjoint bundle of $Im {\Cal R}^+$.
\endproclaim

\noindent{\bf Remark 4.1.9.}  Indeed, in going through the
proof of Proposition 4.1.8 (a) above, one can show that
$$\left \{{\Cal R}^{+}(0_q,X,0)\mid q\in U, X\in L\fg \right\}
 = \bigcup_{q\in U} \{0_q\} \times L^{+}\fg \times \fh.$$
\smallskip

\proclaim
{Proposition 4.1.10} $Im {\Cal R}^{-} = {\Cal I}^{-} \bowtie {\Cal Q},$ 
where
$${\Cal Q} = \bigl \{(0_q,-c(\cdot) Z, Z)\mid q\in U, Z\in \fh\, \bigr \}
\eqno(4.1.13)$$
is a Lie subalgebroid of $Im {\Cal R}^-$ and the ideal ${\Cal I}^{-}$
coincides with the adjoint bundle of $Im {\Cal R}^{-}$ and admits the
following characterization:
$$(0_q,X,0)\in {\Cal I}^{-}_q \,\, \hbox{if and only if}$$
\noindent (a) $X$ is holomorphic in $\Bbb C \setminus \pi \Bbb Z$ with
singularities at the points of the rank one lattice $\pi\Bbb Z$,
\newline
\noindent (b) $X(z)$ is periodic with period $\pi$,
\newline
\noindent (c) $\Pi_{\fh} X_{-1} = 0,$
\newline
\noindent (d) $X$ is bounded as $z\to \infty$ in a
period strip with
$$\lim_{y\to\infty} X(x + iy) = \iota Z- \sum_{\alpha\in <\pi^{\prime}>} 
  (i-c(\alpha(q)))(X_{-1})_{\alpha} e_{\alpha} 
  -2i \sum_{\alpha \in {\overline \pi^{\prime}}^+} (X_{-1})_{\alpha} 
  e_{\alpha},$$
$$\lim_{y\to\infty} X(x - iy) = \iota Z+ \sum_{\alpha\in <\pi^{\prime}>} 
  (i+c(\alpha(q)))(X_{-1})_{\alpha} e_{\alpha} 
  +2i \sum_{\alpha \in {\overline \pi^{\prime}}^-} (X_{-1})_{\alpha} 
  e_{\alpha},$$
for some $Z\in\fh$.  Consequently, 
$X(\pm i \infty) \in \frak{p}^{\pm}_{\pi^{\prime}}$ and 
$$\Pi^{-}_{\fg_{\pi^{\prime}}} X(-i\infty) =
  Ad_{e^{2iq}}\Pi^{+}_{\fg_{\pi^{\prime}}} X(i\infty).\eqno(4.1.14)$$ 
\endproclaim
 
\demo
{Proof}  From the definition of ${\Cal I}^{-}$ , we have
$$\aligned
         & (0_q,X,0) \in {\Cal I}^{-} \\
    \iff & {\Cal R}^{+}(0_q,X,Z) = 0 \,\,\hbox {for some}\,\, Z \in \fh \\
    \iff & \Pi_{\fh} X_{-1} =0, \,\, -\iota Z + R^{+}(q)X =0 \,\,
           \hbox {for some}\,\,\, Z \in \fh \\
    \iff & \Pi_{\fh} X_{-1}=0, \,\, X(z) = \iota Z -(R^{-}(q)X+c(\cdot)
           \Pi_{\fh} X_{-1})(z) \,\,\hbox{for some}\,\, Z\in \fh.
\endaligned $$
Therefore, by Lemma 4.1.7 above and the relation
$c(\alpha(q)) + i = e^{2i\alpha(q)}(c(\alpha(q)) - i)$, we conclude that
$X$ satisfies the properties in (a)-(d).
Conversely, suppose $X\in L\fg$ satisfies the properties in (a)-(d).
Consider 
$$D(z) =\, X(z)+ (R^{-}(q)X)(z).$$
Then by the properties of $X$ and Lemma 4.1.7, $D(z+\pi)=D(z)$ and the
principal part of $D$ at $z=0$ is zero.  Therefore, $D$ extends to a
holomorphic map from $\Bbb C$ to $\fg$.  Moreover, $D$ is bounded
as $z\to\infty$ in the period strip and $\lim_{y\to\infty}D(x\pm iy)=\iota Z$.
Write
$D(z) = \sum_{j} d_{j}(z)x_{j} + \sum_{\alpha\in \Delta} d_{\alpha}(z)
e_{\alpha}.$
Then $d_{j}$ and $d_{\alpha}$ are entire functions for 
$1\leq j \leq N, \alpha\in \Delta$ and are periodic with
period $\pi$. Therefore, when we combine this with the boundedness of 
$d_{j}$ and $d_{\alpha}$ as $z\to\infty$ in the period strip, we conclude
that $d_{j}(z) =d_{j} (=\hbox{constant})$ for each $j$ and 
$d_{\alpha}(z) =d_{\alpha} (=\hbox{constant})$ for each $\alpha$.
But now it follows from 
$\lim_{y\to\infty}D(x\pm iy)=\iota Z$
that we must have
$D(z)=\sum_{j} d_{j}x_{j} =\iota Z$.  Consequently,
$X=\iota Z -R^{-}(q)X$ and this in turn implies that
$- \iota Z + R^{+}(q)X =0$.  As $\Pi_{\fh} X_{-1}=0$, we have 
$(0_q,X,0)\in {\Cal I}^{-}_{q}$, as was to be proved.
The proof of the assertion $Im {\Cal R}^{-} = {\Cal I}^{-} \bowtie {\Cal Q}$
is similar to the one of Proposition 5.1.9 and so we will omit the details.
\pf
\enddemo

\medskip
\subhead
4.2. \ Solution of the integrable trigonometric spin Calogero-Moser\newline 
\phantom{fakke}systems
\endsubhead
\medskip

In principle, we have to solve the factorization problem
$$exp \{\, t(0,0,M(q^0,p^0,\xi^0))\}(q^0)
      =\,\gamma_{+} (t)\,\gamma_{-} (t)^{-1}\eqno(4.2.1)$$
for 
$(\gamma_{+}(t),\gamma_{-}(t))=
((q^0, k_{+} (t), q(t)), (q^0, k_{-} (t), q(t))) \in
Im ({\Cal R}^{+}, {\Cal R}^{-})$
satisfying the condition 
$$\left(T_{\gamma_{+} (t)} {\bbl}_{{\gamma_{+}(t)}^{-1}} {\dot\gamma_{+}(t)},
T_{\gamma_{-} (t)}{\bbl}_{{\gamma_{-}(t)}^{-1}}{\dot \gamma_{-}(t)}\right)
\in \, 
({\Cal R}^+,{\Cal R}^-) (\{q(t)\}\times\{0\}\times L\fg),
\eqno(4.2.2)$$ where
$(q^0,p^0,\xi^0) \in J^{-1}(0) = TU \times ({\Cal U} \cap \fh^{\perp})$
is the initial value of $(q,p,\xi)$ and 
$M(q^0,p^0,\xi^0)(z) = L(q^0,p^0,\xi^0)(z)/z.$
(We shall denote $k_{\pm}(t)(z)$
by $k_{\pm}(z,t)$.)  However, as we shall see in the next two
propositions and their corollary, it actually suffices to
solve for $q(t)$, $k_{+}(0,t)$ and $k_{-}(\pm i\infty,t)$
and we will find the factorization problems for
these quantities from (4.2.1) and (4.2.2) above. 
In what follows, we shall denote by $(q(t),p(t),\xi(t))$
the Hamiltonian flow on $J^{-1}(0)$ generated by ${\Cal H}$
with initial condition $(q(0),p(0),\xi(0)) = (q^0,p^0,\xi^0)$.

\proclaim
{Proposition 4.2.1} With the notations introduced above,
\smallskip
\noindent (a) $L(q(t),p(t),\xi(t))(\pm i\infty)$ exist. Explicitly, 
$$\eqalign{&L(q(t),p(t),\xi(t))(\pm i \infty)\cr
        =\,&L(q(t),p(t),\xi(t))(\pi/2) \mp i\xi(t)\cr
 =\,&p(t) + \sum_{\alpha\in <\pi^{\prime}>}(c(\alpha(q(t)))\mp i)\xi_{\alpha}(t)e_{\alpha}
     \mp 2i\sum_{\alpha\in{\overline \pi^{\prime}}^{\pm}}\xi_{\alpha}(t) e_{\alpha}\cr}
    \eqno(4.2.3)$$
and therefore 
$L(q(t),p(t),\xi(t))(\pm i \infty)\in \frak{p}^{\pm}_{\pi^{\prime}}.$
\smallskip
\noindent (b) $L(q(t),p(t),\xi(t))(\pm i\infty)$ satisfy the Lax equations
$$\eqalign{
         &{d\over dt} L(q(t),p(t),\xi(t))(\pm i \infty)\cr
 = \,&\left[L(q(t),p(t),\xi(t))(\pm i \infty),\, -\sum_{\alpha\in <\pi^{\prime}>}
   \csc^{2} (\alpha(q(t)))\xi_{\alpha}(t)e_{\alpha}\right].\cr}\eqno(4.2.4)$$
\endproclaim

\demo
{Proof} (a) The existence of  $L(q(t),p(t),\xi(t))(\pm i\infty)$ and 
their explicit formulas are obtained from (4.1.4) by noting that
$\lim_{y\to \pm\infty} \cot(iy) = \mp i.$  
\newline
(b) According to Proposition 2.3.5, we have
$$\dot L (q(t),p(t),\xi(t))(z) = [\, L(q(t),p(t),\xi(t))(z), 
(R^{-}(q(t)) M(q(t),p(t),\xi(t)))(z) \,]$$
where
$$\aligned
         & (R^{-}(q(t))M(q(t),p(t),\xi(t)))(z) \\
       = \,&  -c(z)p(t) + \sum_{\alpha\in\Delta
            \setminus <\pi^{\prime}>} \phi_{\alpha}(q(t),z) c(z) \xi_{\alpha}(t)
             e_{\alpha}\\
           & +\sum_{\alpha\in <\pi^{\prime}>}
             \phi_{\alpha}(q(t),z)(c(\alpha(q(t)))+c(z)-c(\alpha(q(t))+z))
             \xi_{\alpha}(t)e_{\alpha}\\
\endaligned
$$ 
by (4.1.8).  Now, it is easy to see from (4.1.3) that
$$
\phi_{\alpha}(q(t),i\infty)=\cases -(-i + \cot \alpha(q(t))), 
  & \alpha\in <\pi^{\prime}> \\
  2i, & \alpha\in
  {\overline \pi^{\prime}}^+ \\
  0, & \alpha\in
  {\overline \pi^{\prime}}^- \endcases 
$$
whereas
$$
\phi_{\alpha}(q(t),-i\infty)=\cases -(i + \cot \alpha(q(t))), 
  & \alpha\in <\pi^{\prime}> \\
  0, & \alpha\in
  {\overline \pi^{\prime}}^+ \\
  -2i, & \alpha\in
  {\overline \pi^{\prime}}^- .\endcases 
$$
Therefore, upon taking the limit as $z = iy \to \pm i\infty$ in the
above expression for  $(R^{-}(q(t))M(q(t),p(t),\xi(t)))(z),$
we find that
$$\aligned
         &(R^{-}(q(t))M(q(t),p(t),\xi(t)))(\pm i\infty)\\
  = \,\, &\pm i L(q(t),p(t),\xi(t))(\pm i\infty)
   -\sum_{\alpha\in <\pi^{\prime}>}
   \csc^{2} (\alpha(q(t)))\xi_{\alpha}(t)e_{\alpha}\\
\endaligned
$$
from which the assertion follows.
\pf
\enddemo

\noindent{\bf Remark 4.2.2.} Although $L(q,p,\xi)(\pm i \infty)$ exist
and satisfy Lax equations, however, they are deficient in the sense
that they do not provide enough conserved quantities for complete
integrability.  In order to establish Liouville integrability,
we must use the Lax operator with spectral parameter
$L(q,p,\xi)(z)$.
\medskip  

We next spell out some of the consequences of the condition in (4.2.2)
which will clarify the relation between
the term 
$-\sum_{\alpha\in <\pi^{\prime}>}
   \csc^{2} (\alpha(q(t)))\xi_{\alpha}(t)e_{\alpha}$
which appears in the Lax equations above for
$L(q(t),p(t),\xi(t))(\pm i\infty)$ and the factors
$k_{\pm}(z,t).$ 

\proclaim
{Proposition 4.2.3} (a) $k_{+}(0,t)\in G_{\pi^{\prime}}$ and satisfies the
equation
$$T_{k_{+}(0,t)}l_{k_{+}(0,t)^{-1}}\dot k_{+}(0,t) =
 -\sum_{\alpha\in <\pi^{\prime}>}\csc^{2} (\alpha(q(t)))\xi_{\alpha}(t)e_{\alpha}.$$
\smallskip
\noindent (b) $k_{-}(\pm i \infty,t)\in P^{\pm}_{\pi^{\prime}}$ and satisfy the
equations
$$\aligned
       &T_{k_{-}(\pm i \infty,t)}l_{k_{-}(\pm i \infty,t)^{-1}}\dot k_{-}(\pm i \infty,t)\\
    \, &= \pm i  L(q(t),p(t),\xi(t))(\pm i \infty)
     -\sum_{\alpha\in <\pi^{\prime}>}\csc^{2} (\alpha(q(t)))\xi_{\alpha}(t)e_{\alpha}.\\
\endaligned
$$
\endproclaim

\demo
{Proof} (a)  It follows from (4.2.1) and (4.2.2) that (see the proof of Theorem
2.2.2 in \c{L2})
$$T_{\gamma_{+} (t)} {\bbl}_{{\gamma_{+}(t)}^{-1}} {\dot\gamma_{+}(t)}
= {\Cal R}^{+}(q(t),0, M(q(t),p(t),\xi(t))).$$
Consequently, we have
$$\aligned
         & T_{k_{+}(z,t)}l_{{k_{+}(z,t)}^{-1}} {\dot k_{+}(z,t)}\\
    =\,\,& (R^{+}(q(t))M(q(t),p(t),\xi(t)))(z)\\
   = \,\,&  M(q(t),p(t),\xi(t))(z) -c(z)p(t) + \sum_{\alpha\in\Delta
            \setminus <\pi^{\prime}>} \phi_{\alpha}(q(t),z) c(z) \xi_{\alpha}(t)
             e_{\alpha}\\
           & +\sum_{\alpha\in <\pi^{\prime}>}
             \phi_{\alpha}(q(t),z)(c(\alpha(q(t)))+c(z)-c(\alpha(q(t))+z))
             \xi_{\alpha}(t)e_{\alpha}.\qquad (*)\\
\endaligned
$$
Since $\cot z = {1\over z} + O(z^3)$ in a deleted neighborhood of $0$,
the $z^0$ term in the Laurent series expansion about $0$ of 
$M(q(t),p(t),\xi(t))(z)$, $c(z)p(t)$ and $\phi_{\alpha}(q(t),z) c(z)$
for $\alpha\in\Delta\setminus <\pi^{\prime}>$ is equal to zero in each case.
On the other hand, for $\alpha \in <\pi^{\prime}>$, the
$z^0$ term in the Laurent series expansion of
$ \phi_{\alpha}(q(t),z)(c(\alpha(q(t)))+c(z)-c(\alpha(q(t))+z))$
about $0$ is $-\csc^{2} (\alpha(q(t)))$.  The formula for
$T_{k_{+}(0,t)}l_{k_{+}(0,t)^{-1}}\dot k_{+}(0,t)$  thus
follows when we let $z\to 0$ in (*) above.
\newline
(b) It also follows from (4.2.1) and (4.2.2) that (see
the proof of Theorem 2.2.2 of \c{L2})
$$T_{\gamma_{-} (t)} {\bbl}_{{\gamma_{-}(t)}^{-1}} {\dot\gamma_{-}(t)}
= {\Cal R}^{-}(q(t),0, M(q(t),p(t),\xi(t)))$$
and hence
$$\aligned
         & T_{k_{-}(z,t)}l_{{k_{-}(z,t)}^{-1}} {\dot k_{-}(z,t)}\\
    =\,\,& (R^{-}(q(t))M(q(t),p(t),\xi(t)))(z).\\
\endaligned
$$
The formulas for 
$T_{k_{-}(\pm i \infty,t)}l_{k_{-}(\pm i \infty,t)^{-1}}\dot k_{-}(\pm i \infty,t)$
then follow from the proof of Proposition 4.2.1 (b).  Finally the
assertion that 
$k_{-}(\pm i \infty,t)\in P^{\pm}_{\pi^{\prime}}$ 
is a consequence of these formulas and Proposition 4.2.1 (a).
\pf
\enddemo

\proclaim
{Corollary 4.2.4} In terms of $k_{+}(0,t)$, we have
$$L(q(t),p(t),\xi(t))(\pm i \infty)= Ad_{k_{+}(0,t)^{-1}}L(q^0,p^0,\xi^0)
(\pm i\infty).$$  Consequently,
$$L(q(t),p(t),\xi(t))(z) = Ad_{k_{+}(0,t)^{-1}}L(q^0,p^0,\xi^0)(z).$$
\endproclaim

\demo
{Proof} By using Proposition 4.2.3 (a) and Proposition 4.2.1 (b),
we can check that $Ad_{k_{+}(0,t)}L(q(t),p(t),\xi(t))(\pm i \infty)$
are constants, hence
$$Ad_{k_{+}(0,t)}L(q(t),p(t),\xi(t))(\pm i \infty) = L(q^0,p^0,\xi^0)(\pm i\infty).
$$
Now it is clear from (4.2.3) that
$$2L(q(t),p(t),\xi(t))(\pi/ 2)= L(q(t),p(t),\xi(t))(i\infty)
  + L(q(t),p(t),\xi(t))(-i\infty)$$
and 
$$-2i\xi(t) = L(q(t),p(t),\xi(t))(i\infty) - L(q(t),p(t),\xi(t))(-i\infty).$$
As 
$$L(q(t),p(t),\xi(t))(z) = L(q(t),p(t),\xi(t))(\pi/ 2) + c(z)\xi(t),$$
the second assertion is a consequence of the first one by virtue
of the above relations.
\pf
\enddemo

Combining the formulas in Proposition 4.2.3 (a) and (b), and the fact
that 
$$L(q(t),p(t),\xi(t))(\pm i \infty) = Ad_{k_{-}(\pm i \infty,t)^{-1}}L(q^0,p^0,\xi^0)
(\pm i \infty),\eqno(4.2.5)$$ 
we  obtain the following factorization problems on $P^{\pm}_{\pi^{\prime}}$\,:
$$e^{itL(q^0,p^0,\xi^0)(i \infty)} = k_{-}(i\infty, t)k_{+}(0,t)^{-1}, \eqno(4.2.6)$$
$$e^{-itL(q^0,p^0,\xi^0)(-i \infty)} = k_{-}(-i\infty, t)k_{+}(0,t)^{-1} \eqno(4.2.7)$$
where $k_{+}(0,t)$ and $k_{-}(\pm i\infty)$ are to be determined.
The nature of these factorization problems are of course
quite different from that of those
in the well-known group-theoretic scheme
for constant r-matrices.(Compare, for example, the factorization
problems in \c{RSTS},\c{STS}, \c{DLT} with our solution of
(4.2.6), (4.2.7) below.)

We shall use the following notation: for  $g^{\pm}\in  P^{\pm}_{\pi^{\prime}}$,
$\bn^{\pm}(g^{\pm})\in N^{\pm}_{\pi^{\prime}}$, $\bla^{\pm}(g^{\pm})\in G_{\pi^{\prime}}$ 
will denote the factors in the unique factorization
$g^{\pm} = \bn^{\pm}(g^{\pm}) \bla^{\pm}(g^{\pm})$.
In order to solve (4.2.6) and (4.2.7), note that
$$(q^0, k_{-} (t), q(t)) = (q^0,\widehat{k}_{-}(t),q^0)(q^0,
   e^{c(\cdot)(q^0-q(t))},q(t))\eqno(4.2.8)$$
by the global version of Proposition 4.1.10 where
 $(q^0,\widehat{k}_{-}(t),q^0)$ is in the Lie group bundle
integrating ${\Cal I}^{-}$ and the second factor 
$(q^0,e^{c(\cdot)(q^0-q(t))},q(t))$
is in the Lie groupoid integrating ${\Cal Q}$. 
Consequently, the factorization problems on $P^{\pm}_{\pi^{\prime}}$ in
(4.2.6) and (4.2.7) can be recast in the form
$$e^{itL(q^0,p^0,\xi^0)(i \infty)} = \bn^{+}(\widehat{k}_{-}(i\infty, t))
\bla^{+}(\widehat{k}_{-}(i\infty, t))e^{-i(q^0-q(t))}k_{+}(0,t)^{-1}, \eqno(4.2.9)$$
$$e^{-itL(q^0,p^0,\xi^0)(-i \infty)} =\bn^{-}(\widehat{k}_{-}(-i\infty, t))
\bla^{-}(\widehat{k}_{-}(-i\infty, t))e^{i(q^0-q(t))}k_{+}(0,t)^{-1}. \eqno(4.2.10)$$
Now, from the fact that 
$e^{itL(q^0,p^0,\xi^0)(i \infty)}\in P^{+}_{\pi^{\prime}}$, we can find unique
$n_{+}(t)\in N^{+}_{\pi^{\prime}}$, $g_{+}(t)\in G_{\pi^{\prime}}$ 
satisfying $n_{+}(0)=g_{+}(0)=1$ such that 
$$e^{itL(q^0,p^0,\xi^0)(i \infty)} = n_{+}(t) g_{+}(t).\eqno(4.2.11)$$
Similarly, we can find unique 
$n_{-}(t )\in N^{-}_{\pi^{\prime}}$, $g_{-}(t)\in G_{\pi^{\prime}}$ 
satisfying $n_{-}(0)=g_{-}(0)=1$ such that 
$$e^{-itL(q^0,p^0,\xi^0)(i \infty)} = n_{-}(t) g_{-}(t).\eqno(4.2.12)$$
By comparing (4.2.9) (resp.~ (4.2.10))with (4.2.11) (resp.~(4.2.12)), we
obtain
$$\bn^{+}(\widehat{k}_{-}(i\infty,t)) = n_{+}(t),\,\,
  \bn^{-}(\widehat{k}_{-}(-i\infty,t)) = n_{-}(t).\eqno(4.2.13)$$
Hence the factorization problems reduce to
$$g_{+}(t) =\bla^{+}(\widehat{k}_{-}(i\infty, t))e^{-i(q^0-q(t))}k_{+}(0,t)^{-1},
\eqno(4.2.14)$$
$$g_{-}(t)=\bla^{-}(\widehat{k}_{-}(-i\infty, t))e^{i(q^0-q(t))}k_{+}(0,t)^{-1}. 
\eqno(4.2.15)$$
But from the global version of (4.1.14), we have
$$\bla^{-}(\widehat{k}_{-}(-i\infty, t))= e^{2iq^0}
  \bla^{+}(\widehat{k}_{-}(i\infty, t))e^{-2iq^0}.\eqno(4.2.16)$$
Substitute this into (4.2.15) above, we find
$$e^{-2iq^0}g_{-}(t) = \bla^{+}(\widehat{k}_{-}(i\infty, t))e^{-i(q^0+q(t))}
k_{+}(0,t)^{-1}.\eqno(4.2.17)$$
Consequently, when we eliminate 
$\bla^{+}(\widehat{k}_{-}(i\infty, t))$
from (4.2.14) and (4.2.17), we obtain the following factorization problem
on $G_{\pi^{\prime}}$:
$$g_{-}(t)^{-1}e^{2iq^0}g_{+}(t) = k_{+}(0,t)e^{2iq(t)}k_{+}(0,t)^{-1}.\eqno(4.2.18)$$
But $G_{\pi^{\prime}}$ is a reductive Lie group, hence we can find (for at least
small values of $t$) $x(t)\in G_{\pi^{\prime}}$ (unique to transformations
$x(t)\rightarrow x(t)\delta(t)$ where $\delta(t)\in H$) and unique
$d(t)\in H$ such that
$$g_{-}(t)^{-1}e^{2iq^0}g_{+}(t) = x(t) d(t) x(t)^{-1} \eqno(4.2.19)$$
with $x(0)=1, d(0) = e^{2iq^0}.$  This determines $q(t)$ via
the formula
$$q(t) = {1\over 2i} log\,d(t).\eqno(4.2.20)$$
On the other hand, let us fix one such $x(t)$.  We shall seek
$k_{+}(0,t)$ in the form 
$$k_{+}(0,t) = x(t) h(t), \quad h(t) \in H.\eqno(4.2.21)$$
To determine $h(t)$, we shall impose the following condition 
(which is a corollary of Proposition 4.2.3 (a)):
$$\Pi_{\fh} T_{k_{+}(0,t)} l_{k_{+}(0,t)^{-1}} \dot k_{+}(0,t) =0\eqno(4.2.22)$$
where $\Pi_{\fh}$ is the projection map to $\fh$ relative to
the direct sum decomposition $\fg_{\pi^{\prime}} = \fh + \sum_{\alpha\in
<\pi^{\prime}>} \fg_{\alpha}.$
Substitute (4.2.21) into (4.2.22), we see that $h(t)$ satisfies the
equation
$$\dot h(t) = -  T_{e} l_{h(t)} (\Pi_{\fh} T_{x(t)} l_{{x(t)}^{-1}}\dot x(t))
\eqno(4.2.23)$$
with $h(0)=1.$  Solving the equation explicitly, we find that
$$h(t)=  exp\left\{-\int _{0}^{t}{\Pi_{\fh} (T_{x(\tau)} l_{{x(\tau)}^{-1}} 
\dot x(\tau))}\, d\tau \right\} .\eqno(4.2.24)$$

\proclaim
{Theorem 4.2.5} Let 
$(q^0,p^0,\xi^0) \in J^{-1}(0) = TU \times ({\Cal U} \cap \fh^{\perp}).$
Then the Hamiltonian flow on $J^{-1} (0)$ generated by
$$\eqalign{{\Cal H}(q,p,\xi) = &{1\over 2} \sum_{i} p_{i}^{2} - {1\over 2}
  \sum_{\alpha \in <\pi^{\prime}>} \left(\frac{1}{\sin^{2}\alpha(q)}
  -{1\over 3}\right) {\xi_{\alpha}\xi_{-\alpha}}
  -{5\over 6} \sum_{\alpha\in\Delta\setminus <\pi^{\prime}>}
  {\xi_{\alpha}\xi_{-\alpha}}\cr
   & -{1\over 3} \sum_{i} \xi^{2}_{i}.\cr}$$
with initial condition $(q(0), p(0), \xi(0)) = (q^0,p^0,\xi^0)$ is
given by
$$\eqalign{& q(t) =\, {1\over 2i} log\,d(t), \cr
       & \xi (t) =\, Ad_{k_{+}(0,t)^{-1}} \xi^{0},\cr
       & p(t) = \, Ad_{k_{+}(0,t)^{-1}} L(q^0,p^0,\xi^0)(\pm i \infty)
         - \sum_{\alpha\in <\pi^{\prime}>}(c(\alpha(q(t)))\mp i)\xi_{\alpha}(t)
         e_{\alpha}\cr       
       &\qquad\,\,\, \pm 2i\sum_{\alpha\in{\overline \pi^{\prime}}^{\pm}}\xi_{\alpha}(t) 
        e_{\alpha}\cr}
    \eqno(4.2.25)$$
where $d(t)$ and $k_{+}(0,t)$ are constructed from the above
procedure.
\endproclaim

\demo
{Proof} The formula for $\xi(t)$ is a consequence of Corollary 4.2.4
and the relation
$$-2i\xi(t) = L(q(t),p(t),\xi(t))(i\infty) - L(q(t),p(t),\xi(t))(-i\infty).$$
On the other hand, the formula for $p(t)$ follows by equating the
two different expressions for $L(q(t),p(t),\xi(t))(\pm i\infty)$
in (4.2.3) and in Corollary 4.2.4.
\pf
\enddemo

By Poisson reduction, we can now write down the solution of the
associated integrable model on $TU\times \fg_{red}$ with Hamiltonian
${\Cal H}_{0}$
whose equations of motion are given in Proposition 4.1.3, as in
Corollary 3.2.2.
 
\proclaim
{Corollary 4.2.6} Let $(q^0,p^0,s^0) \in TU \times \fg_{red}$ and
suppose $s^0 =  Ad_{g(\xi^{0})^{-1}} \xi^{0}$ where 
$\xi^{0}\in {\Cal U} \cap \fh^{\perp}.$  Then the Hamiltonian
flow generated by ${\Cal H}_{0}$ with initial condition
$(q(0),p(0),s(0)) = (q^0,p^0,s^0)$ is given by
$$\eqalign{
       & q(t) =\, d(t), \cr
       & s(t) =\, Ad_{\bigl({\widetilde k}_{+}(0,t)\,
                  g \bigl(Ad_{{\widetilde k}_{+}(0,t)^{-1}}\, s^{o}\bigr)
       \bigr)^{-1}} s^{0},\cr
       & p(t) =\, Ad_{\bigl({\widetilde k}_{+}(0,t)\,
                  g\bigl(Ad_{{\widetilde k}_{+}(0,t)^{-1}} s^{o}\bigr)
         \bigr)^{-1}} L(q^0,p^0,s^0)(\pm i\infty)\cr
       &\qquad\,\,\, - \sum_{\alpha\in <\pi^{\prime}>}(c(\alpha(q(t)))\mp i)
         s_{\alpha}(t)e_{\alpha}
         \pm 2i\sum_{\alpha\in{\overline \pi^{\prime}}^{\pm}}
       s_{\alpha}(t) e_{\alpha}\cr}\eqno(4.2.26)$$
where $ {\widetilde k}_{+}(0,t) = g(\xi^{0})^{-1} k_{+}(0,t) g(\xi^{0})$
and $k_{+}(0,t)$, $d(t)$ are as in Theorem 4.2.5.
\endproclaim

\noindent{\bf Remark 4.2.7.} The reader should contrast the
factorization problems in this section with the ones in
\c{L2}. Although the Hamiltonians are rather similar
(we can transform the hyperbolic spin CM systems in
\c{L2} to trigonometric ones), however, the factorization
problems involved are quite different.

\bigskip
\bigskip

\subhead
5.\  The elliptic spin Calogero-Moser systems
\endsubhead
\medskip
\subhead
5.1.\ Lax operators, Hamiltonian equations and the Lie subalgebroids
\endsubhead
\medskip

In this section, $\wp(z)$ is the Weierstrass $\wp$-function with
periods $2 \omega_{1}$,$2\omega_{2} \in \Bbb C$, and
$\sigma (z)$, $\zeta (z)$ are the related Weierstrass sigma-function and
zeta-function.

We consider the following elliptic dynamical r-matrix with spectral
parameter, given by
$$r(q,z) = \zeta(z)\sum_{i} x_{i}\otimes x_{i} -\sum_{\alpha\in \Delta}
  l(\alpha(q),z) e_{\alpha}\otimes e_{-\alpha} \eqno(5.1.1)$$
where
$$l(w,z) = -\frac{\sigma(w+z)}{\sigma(w)\sigma(z)}.\eqno(5.1.2)$$
Then the associated spin Calogero-Moser system on $TU\times\fg$ is 
called the elliptic spin Calogero-Moser system.  Explicitly, the
Hamiltonian is of the form
$${\Cal H}(q,p,\xi)  = {1\over 2} \sum_{i} p_{i}^{2} - {1\over 2}
  \sum_{\alpha \in \Delta} \wp(\alpha(q)) {\xi_{\alpha}\xi_{-\alpha}}
   \eqno(5.1.3)$$
and the Lax operator is given by
$$L(q,p,\xi)(z) =p+\zeta(z)\sum_{i}\xi_{i}x_{i} -
  \sum_{\alpha\in \Delta} l(\alpha(q),z)\xi_{\alpha}e_{\alpha}.
  \eqno(5.1.4)$$

Our next result gives the Hamiltonian equations of motion
generated by ${\Cal H}.$  Using the same method of calculation
as in the proof of Proposition 3.1.4, we can also compute the
corresponding equations generated by its reduction 
$${\Cal H_{0}}(q,p,s)  = {1\over 2} \sum_{i} p_{i}^{2} - {1\over 2}
  \sum_{\alpha \in \Delta} \wp(\alpha(q)) {s_{\alpha}s_{-\alpha}}
   \eqno(5.1.5)$$
on $TU\times \fg_{red}.$

\proclaim
{Proposition 5.1.1} The Hamiltonian equations of motion generated by
${\Cal H}$ on $TU\times \fg$ are given by
$$\eqalign{& \dot q = p, \cr
           & \dot p = {1\over 2} \sum_{\alpha \in \Delta}
             \wp^{\prime}(\alpha(q))\xi_{\alpha} \xi_{-\alpha} H_{\alpha},\cr
           & \dot \xi = \left[\,\xi,- \sum_{\alpha \in \Delta}
             \wp(\alpha(q))\xi_{\alpha} 
             e_{\alpha}\,\right ].\cr}\eqno(5.1.6)$$
\endproclaim

\proclaim
{Proposition 5.1.2} The Hamiltonian equations of motion generated by
${\Cal H}_0$ on the reduced Poisson manifold $TU\times \fg_{red}$
are given by
$$\aligned
         & \dot q = p, \\
         & \dot p = {1\over 2} \sum_{\alpha \in \Delta}
             \wp^{\prime}(\alpha(q))s_{\alpha}s_{-\alpha} H_{\alpha},\cr \\
         & \dot s = [\,s, {\Cal M}\,] 
\endaligned
$$
where 
$${\Cal M} =  - \sum_{\alpha \in \Delta}
           \wp(\alpha(q)) s_{\alpha} e_{\alpha}
          + \sum_{i,j} C_{ji}
           \sum \Sb \alpha \in \Delta \\
           \alpha_{j}-\alpha \in \Delta \endSb N_{\alpha, \alpha_{j}-\alpha}\, 
           \wp(\alpha(q))s_{\alpha} s_{\alpha_{j}-\alpha} h_{\alpha_i}.$$
(Here the notation $N_{\alpha,\beta}$ is as in Proposition 3.1.4.)
\endproclaim

\proclaim
{Proposition 5.1.3} The classical dynamical r-matrix $R$ associated
with the elliptic dynamical r-matrix with spectral parameter in (5.1.1) 
is given by
$$\eqalign{(R(q)X)(z) & = {1\over 2}X(z) +\sum_{k=0}^{\infty} \frac{
           \zeta^{(k)}(-z)}{k!}\, \Pi_{\fh} X_{-(k+1)}\cr
           & +\sum_{k=0}^{\infty} \frac{1}{k!}\sum_{\alpha\in \Delta}
           {d^{k}\over dw^{k}}{\Big|_{w=0}}\, l(\alpha(q),z-w)
           (X_{-(k+1)})_{\alpha}
           e_{\alpha}.\cr}\eqno(5.1.7)$$
\endproclaim

\demo
{Proof} Here, we have used the formula $l(-w,z) = -l(w,-z).$  Otherwise,
the proof is similar to that of Proposition 4.1.4.
\pf
\enddemo

\proclaim
{Corollary 5.1.4} On $J^{-1}(0)$, we have 
$$\eqalign{& (R(q)M(q,p,\xi))(z) \cr
         = &{1\over 2} M(q,p,\xi)(z) -\zeta(z)p +\sum_{\alpha\in \Delta}
             l(\alpha(q),z)(\zeta(\alpha(q))+\zeta(z)-\zeta(\alpha(q)+z))
             \xi_{\alpha} e_{\alpha}\cr}\eqno(5.1.8)$$
where $M(q,p,\xi)(z) = L(q,p,\xi)(z)/z.$
\endproclaim

\demo
{Proof} In a deleted neighborhood of $0$, we have the expansion
$$M(q,p,\xi)(z) = \frac{\xi}{z^2} + \frac{1}{z}\left(p +
  \sum_{\alpha} \zeta(\alpha(q))\xi_{\alpha}e_{\alpha}\right)
  + O(1).$$
On the other hand, by direct differentiation, we find
$${d\over dw}{\Big|_{w=0}} l(\alpha(q),z-w)
  = -l(\alpha(q),z)(\zeta(\alpha(q)+z) -\zeta(z)).$$
Therefore, on using (5.1.7), we obtain the desired formula.
\pf
\enddemo

\noindent {\bf Remark 5.1.5.} Using (5.1.8), we can check that in this
case, the equations in (5.1.6) can be recovered from the Lax equation
$\dot L(q,p,\xi) = [L(q,p,\xi), R(q)M(q,p,\xi)]$ on $J^{-1}(0).$
The computation makes use of the following identities:\newline
\noindent (i) $l(w,z)l(-w,z) = \wp(z) -\wp(w)$,\newline
\noindent (ii) $ l(x,z)l(y,z) [\zeta(x+z)-\zeta(x)-\zeta(y+z)+\zeta(y)]
     = l(x+y,z)[\wp(x)-\wp(y)]$,\newline
\noindent (iii) $\zeta(x+y) -\zeta(x)-\zeta(y) ={1\over 2}\frac{
     \wp^{\prime}(x)-\wp^{\prime}(y)} {\wp(x)-\wp(y)}.$\newline
We shall leave the details to the interested reader.

Our next lemma is a simple consequence of the fact that
for $k\geq 0$, we have
$\zeta^{(k)}(-z) = -k!z^{-(k+1)} + O(1)$,
${d^{k}\over dw^{k}}{\Big|_{w=0}} l(\alpha(q),z-w) = -k!z^{-(k+1)} +O(1)$
in a deleted neighborhood of $0$.  
\proclaim
{Lemma 5.1.6} For $X\in L\fg$, $R^{+}(q)X\in L^{+}\fg$.
\endproclaim

\proclaim
{Lemma 5.1.7} For $X\in L\fg$, $R^{-}(q)X + \zeta(\cdot)\Pi_{\fh}X_{-1}$
has singularities at the points of the rank $2$ lattice
$$\Lambda = 2\omega_{1} \Bbb Z + 2\omega_{2} \Bbb Z\eqno(5.1.9)$$
and is holomorphic in $\Bbb C \backslash \Lambda.$  
Moreover, the quasi-periodicity condition
$$(R^{-}(q)X+\zeta(\cdot)\Pi_{\fh}X_{-1})(z+2\omega_{i})=
  Ad_{e^{2\eta_{i} q}} (R^{-}(q)X+\zeta(\cdot)\Pi_{\fh}X_{-1})(z)\eqno(5.1.10)$$
holds, where $\eta_{i} =\zeta(\omega_{i}),$ $i=1,2$.
\endproclaim

\demo
{Proof} From (5.1.7), we obtain
$$\aligned
         &(R^{-}(q)X + \zeta(\cdot) \Pi_{\fh}\,X_{-1})(z) \\
       = &\sum_{k=1}^{\infty} \frac{
           \zeta^{(k)}(-z)}{k!} \, \Pi_{\fh} X_{-(k+1)}
          +\sum_{k=0}^{\infty} \sum_{\alpha\in \Delta}\frac{1}{k!}
           {d^{k}\over dw^{k}}{\Big|_{w=0}} l(\alpha(q),z-w) 
           (X_{-(k+1)})_{\alpha} e_{\alpha}
\endaligned $$
from which it is clear that 
$R^{-}(q)X + \zeta(\cdot) \Pi_{\fh}\,X_{-1}$ is holomorphic in
$\Bbb C \backslash \Lambda$ with singularities at the points of $\Lambda$.   
On the other hand,
it is easy to check that
$${d^{k}\over dw^{k}}{\Big|_{w=0}} l(\alpha(q),z+2\omega_{i}-w) 
  =e^{2\eta_{i}\alpha(q)} {d^{k}\over dw^{k}}{\Big|_{w=0}} l(\alpha(q),z-w),
  \quad k\geq 0.$$ 
Hence the second assertion follows.
\pf
\enddemo

The proof of our next proposition is obvious.
\proclaim
{Proposition 5.1.8} (a) $Im {\Cal R}^{+} = \bigcup_{q\in U} \{0_q\} \times
L^{+}\fg \times \fh$.
\smallskip
\noindent (b) ${\Cal I}^{+} = \bigcup_{q\in U} \{0_q\} \times L^{+}\fg \times 
\{0\}$ = adjoint bundle of $Im {\Cal R}^+$.
\endproclaim

Indeed, we can show more, namely, 
$$\left \{{\Cal R}^{+}(0_q,X,0)\mid q\in U, X\in L\fg \right\}
 = \bigcup_{q\in U} \{0_q\} \times L^{+}\fg \times \fh.\eqno(5.1.11)$$

\proclaim
{Proposition 5.1.9} $Im {\Cal R}^{-} = {\Cal I}^{-} \bowtie {\Cal Q},$ 
where
$${\Cal Q} = \bigl \{(0_q,-\zeta(\cdot) Z, Z)\mid q\in U, Z\in \fh\, \bigr \}
\eqno(5.1.12)$$
is a Lie subalgebroid of $Im {\Cal R}^-$ and the ideal ${\Cal I}^{-}$
coincides with the adjoint bundle of $Im {\Cal R}^{-}$ and admits the
following characterization:
$$(0_q,X,0)\in {\Cal I}^{-}_q \,\, \hbox{if and only if}$$
\noindent (a) $X$ is holomorphic in $\Bbb C \backslash \Lambda$ with
singularities at the points of $\Lambda$,
\newline
\noindent (b) $X(z+2\omega_i) = Ad_{e^{2\eta_{i}q}} X(z), \,\, i=1,2$,
\newline
\noindent (c) $\Pi_{\fh} X_{-1} = 0.$
\endproclaim

\demo
{Proof} As in the proof of Proposition 4.1.10, we have
$$\aligned
         & (0_q,X,0) \in {\Cal I}^{-} \\
    \iff & \Pi_{\fh} X_{-1}=0, \,\, X(z) = \iota Z -(R^{-}(q)X+\zeta(\cdot)
           \Pi_{\fh} X_{-1})(z) \,\,\hbox{for some}\,\, Z\in \fh.
\endaligned $$
Therefore, it follows from Lemma 5.1.7 that $X$ satisfies the properties
in (a)-(c).  Conversely, suppose $X\in L\fg$ satisfies the properties
in (a)-(c).  Consider
$$D(z) = X(z) + (R^{-}(q) X)(z).$$
Then $D(z)$ satisfies the quasi-periodicity condition
$D(z+2\omega_{i})=Ad_{e^{2\eta_{i}q}} D(z), i=1,2$
and the principal part of $D$ at $0$ is zero.  Hence $D$ extends to
a holomorphic map from $\Bbb C$ to $\fg.$    Write
$D(z) = \sum_{j} d_{j}(z)x_{j} + \sum_{\alpha\in \Delta} d_{\alpha}(z)
e_{\alpha}.$
Then $d_{j}$ and $d_{\alpha}$ are entire functions for 
$1\leq j \leq N, \alpha\in \Delta$
and the quasi-periodicity condition implies that for $i=1,2$, we have
$$d_{j}(z+2\omega_{i}) = d_{j}(z), \quad 1\leq j\leq N,\eqno(5.1.13)$$
$$d_{\alpha}(z+2\omega_{i}) = d_{\alpha}(z) e^{2\eta_{i}\alpha(q)},\quad
    \alpha\in \Delta.\eqno(5.1.14)$$
From (5.1.13) and Liouville's theorem, it follows that 
$d_{j}(z) = d_{j} (= \hbox{constant})$ for
each $j$.  On the other hand, observe that
the meromorphic function 
$l(\alpha(q),z)$  satisfies the same quasi-periodicity condition as
$d_{\alpha}$, that is,
$l(\alpha(q), z+2\omega_{i}) = l(\alpha(q),z) e^{2\eta_{i}\alpha(q)}.$
Hence we conclude from (5.1.14) and the above observation that
$d_{\alpha}(z) = f_{\alpha} (z) l(\alpha(q),z)$, where 
$f_{\alpha}$ is an elliptic function.  But the order of a non-constant 
elliptic function is never less than $2$.  Therefore, as $l(\alpha(q),z)$
has simple poles at the points of $\Lambda$, we must have
$f_{\alpha}\equiv d_{\alpha} \equiv 0$ for each $\alpha\in \Delta.$  
Hence we have shown
that $X= \iota Z-R^{-}(q)X$, where $Z=\sum_{j} d_{j}x_{j}.$  This 
in turn implies that $-\iota Z+R^{+}(q)X=0$ and so
$(0_q,X,0)\in {\Cal I}^{-}.$

Next, we show 
$Im {\Cal R}^{-} = {\Cal I}^{-} \oplus {\Cal Q}$.
Consider an arbitrary element ${\Cal R}^{-}(0_q,X,Z)$ in $Im {\Cal R}^-$.  
Clearly, we have  the
decomposition
$$\aligned
         & {\Cal R}^{-}(0_q,X,Z) \\
     =\, & (0_q, -\iota Z + R^{-}(q)X+\zeta(\cdot)\Pi_{\fh}X_{-1},0)\\
         & + (0_q, -\zeta(\cdot) {\Pi_{\fh} X_{-1}},\Pi_{\fh} X_{-1})\\
\endaligned $$
where the first term is in ${\Cal I}^-$ (by Lemma 5.1.7 and the characterization
of ${\Cal I}^-$ which we established above) and the second term 
is in
${\Cal Q}$.  This shows that 
$Im {\Cal R
}^{-} \subset {\Cal I}^{-} \oplus {\Cal Q}.$  Conversely,
take an arbitrary element 
$(0_q,X,0)+(0_q,-\zeta(\cdot)Z,Z)\in {\Cal I}^{-} \oplus {\Cal Q}$.
From the definition of ${\Cal I}^-$, we have
${\Cal R}^{+}(0_q,X,Z^{\prime})=0$ for some $Z^{\prime}\in \fh$.  Let
$Y=-X +\zeta(\cdot)Z.$  Then 
$$\aligned
         &{\Cal R}^{-}(0_q,Y, -Z^{\prime}) \\
     =\, & (0_q, \iota Z^{\prime}-R^{-}(q)X+R^{-}(q)\zeta(\cdot)Z, Z)
          \quad (\because \,\, \Pi_{\fh} Y_{-1}=Z)\\
     =\, & (0_q,\iota Z^{\prime} -R^{+}(q)X +X+ R^{-}(q)\zeta(\cdot)Z,Z)\\
     =\, & (0_q, X, 0) + (0_q,-\zeta(\cdot)Z, Z) \quad 
           (\because \,\, \iota Z^{\prime} -R^{+}(q)X=0
            \,\, and \,\,R^{-}(q)\zeta(\cdot)Z = -\zeta(\cdot)Z)\\
\endaligned $$
and this establishes the reverse inclusion 
 ${\Cal I}^{-} \oplus {\Cal Q}\subset Im {\Cal R}^{-}$.
The assertion that ${\Cal I}^-$ coincides with the adjoint bundle
of $Im {\Cal R}^-$ is now clear.
\pf
\enddemo
\smallskip

From Proposition 5.1.8, it follows that
$$\eqalign{Im {\Cal R}^{+}/{\Cal I}^{+}& = \bigcup_{q\in U} 
  \lbrace 0_q \rbrace \times (L^{+}\fg/L^{+}\fg)\times \fh\cr
  & \simeq \bigcup_{q\in U} \lbrace 0_q \rbrace \times
    \lbrace 0 \rbrace \times \fh\cr}\eqno(5.1.15)
$$
where the identification map is given by
$$(0_q, X+ L^{+}\fg, Z) \mapsto (0_q,0,Z).\eqno(5.1.16)$$
Similarly, as a consequence of Proposition 5.1.9, we obtain
$$Im {\Cal R}^{-}/{\Cal I}^{-} \simeq {\Cal Q}.\eqno(5.1.17)$$
This time, the identification is given by the map
$$(0_q,X,Z) + {\Cal I}^{-}_q \mapsto (0_q, -\zeta(\cdot) Z, Z).\eqno(5.1.18)$$
The following proposition is now obvious.

\proclaim
{Proposition 5.1.10} The isomorphism $\theta : Im {\Cal R}^{+}/{\Cal I}^{+}
\longrightarrow Im {\Cal R}^{-}/{\Cal I}^{-}$ defined in Proposition 2.1.2 (b)
is given by
$$\theta (0_q,0,Z) = (0_q,-\zeta(\cdot) Z, Z).$$
\endproclaim
\medskip

\subhead
5.2.\ Solution of the integrable elliptic spin Calogero-Moser systems
\endsubhead
\medskip

We are now ready to discuss the factorization problem
$$exp \{\, t(0,0,M(q^0,p^0,\xi^0))\}(q^0)
      =\,\gamma_{+} (t)\,\gamma_{-} (t)^{-1}\eqno(5.2.1)$$
where 
$(\gamma_{+}(t),\gamma_{-}(t))=
((q^0, k_{+} (t), q(t)), (q^0, k_{-} (t), q(t))) \in
Im ({\Cal R}^{+}, {\Cal R}^{-})$
is to be determined subject to the constraint in (2.2.4) with
$\fg$ replaced by $L\fg$ and where
$(q^0,p^0,\xi^0) \in J^{-1}(0) = TU \times ({\Cal U} \cap \fh^{\perp})$
is the initial value of $(q,p,\xi)$. (Recall that
$M(q^0,p^0,\xi^0)(z) =L(q^0,p^0,\xi^0)(z)/z.$)  Note that by the global
version of Proposition 5.1.9, we have the unique factorization
$$(q^0, k_{-} (t), q(t)) = (q^0,\widehat{k}_{-}(t),q^0)(q^0,
   e^{\zeta(\cdot)(q^0-q(t))},q(t))\eqno(5.2.2)$$
where $(q^0,\widehat{k}_{-}(t),q^0)$ is in the Lie group bundle
integrating ${\Cal I}^{-}$ and the second factor 
$(q^0,e^{\zeta(\cdot)(q^0-q(t))},q(t))$
is in the Lie groupoid integrating ${\Cal Q}$.  As before,
we denote $k_{\pm}(t)(z)$ by $k_{\pm}(z,t)$.  Also, denote
$\widehat{k}_{-}(t)(z)$ by $\widehat{k}_{-}(z,t)$. Then
$k_{+}(\cdot,t)\in L^{+}G$, while 
$\widehat{k}_{-}(\cdot,t)= k_{-}(\cdot,t) e^{\zeta(\cdot)(q(t)-q^0)}$ enjoys
the following properties:
\smallskip
\noindent (a) $\widehat{k}_{-}(\cdot,t)$ is holomorphic in 
$\Bbb{C}\backslash{\Lambda}$
with singularities at the points of $\Lambda$,\newline
\noindent (b) $\widehat{k}_{-}(z+2\omega_{i}, t) = e^{2\eta_{i}q^0}
    \widehat{k}_{-}(z,t)e^{-2\eta_{i}q^0},\quad i=1,2,$\newline
\noindent (c) $\left({d\over dt}{\Big|_{t=0}}\widehat{k}_{-}(z,t)\right)_{-1}
     \in \fh^{\perp}$.
\smallskip
From the factorization problem on the Lie groupoid above and (5.2.2),
it follows that 
$$e^{t M(q^0,p^0,\xi^0)(z)} = k_{+}(z,t)e^{\zeta(z)(q(t)-q^0)}\widehat{k}_{-}(z,t)^{-1} 
  \eqno(5.2.3)$$
where $q(t)$, $k_{+}(\cdot,t)$ and $\widehat{k}_{-}(\cdot,t)$ are to be
determined.  To do so, let us introduce the following gauge transformations
of $L(q,p,\xi)$ and $M(q,p,\xi)$:
$$\eqalign{&L^{e}(q,p,\xi)(z):= Ad_{e^{-\zeta(z)q}} L(q,p,\xi)(z),\cr
           &M^{e}(q,p,\xi)(z):= Ad_{e^{-\zeta(z)q}} M(q,p,\xi)(z)\cr}\eqno(5.2.4)
$$
for $(q,p,\xi)\in J^{-1}(0) = TU\times ({\Cal U}\cap \fh^{\perp})$.
Then the problem in (5.2.3) can be reformulated in the form
$$e^{t M^{e}(q^0,p^0,\xi^0)(z)}=k^{s}_{+}(z,t)k^{e}_{-}(z,t)^{-1}\eqno(5.2.5)$$
where
$$ k^{s}_{+}(z,t)= e^{-\zeta(z)q^0}k_{+}(z,t)e^{\zeta(z)q(t)}\eqno(5.2.6)$$
and
$$k^{e}_{-}(z,t)=e^{-\zeta(z)q^0}\widehat{k}_{-}(z,t)e^{\zeta(z)q^0}.\eqno(5.2.7)$$
Now, on using the property of $\widehat{k}_{-}(z,t)$ in property (b)
above and the fact that
$l(w,z+2\omega_i) =e^{2\eta_{i}w} l(w,z)$, it is straight forward to check that
$$ k^{e}_{-}(z+ 2\omega_{i},t)=k^{e}_{-}(z,t),\eqno(5.2.8)$$
and
$$L^{e}(q,p,\xi)(z+2\omega_{i}) = L^{e}(q,p,\xi)(z)\eqno(5.2.9)$$
for $i =1,2$.
In view of this, it is natural to introduce the elliptic curve
$\Sigma = \Bbb{C}/\Lambda$ where $\Lambda$ is the rank 2 lattice
in (5.1.9).  Thus we can regard $k^{e}_{-}(\cdot,t)$ as a holomorphic
map on $\Sigma\backslash \{0\}$ taking values in $G$.  On the other
hand, the factor $k^{s}_{+}(\cdot,t)$ in (5.2.6) is holomorphic in
a deleted neighborhood of $0\in \Sigma$.  Hence we can think of
(5.2.5) as a factorization problem on a small circular contour
centered at $0\in \Sigma$ where   $k^{s}_{+}(\cdot,t)$ and 
$k^{e}_{-}(\cdot,t)$ have analyticity properties as indicated
above and satisfying additional constraints.  Indeed, it
follows from (5.2.6) and property (c) above for $\widehat{k}_{-}(\cdot,t)$
that
$$k^{s}_{+}(z,t)\sim e^{-q^0\over z}k_{+}(0,t)e^{q^0\over z}\hbox{\quad in a deleted
  neighborhood of}\quad 0,\eqno(5.2.10)$$
$$\left({d\over dt}{\Big|_{t=0}}\widehat{k}^{e}_{-}(z,t)\right)_{-1}
     \in \fh^{\perp}.\eqno(5.2.11)$$
Note that if $(q(t),p(t),\xi(t))$ is the solution of the Hamiltonian
equations in (5.1.6) satisfying the initial condition
$(q(0),p(0),\xi(0)) =(q^0,p^0,\xi^0)$, then by Theorem 2.2.2 and
our discussion above, we have
$$\eqalign{&L^{e}(q(t),p(t),\xi(t))(z)\cr
           =\, &k^{s}_{+}(z,t)^{-1}L^{e}(q^0,p^0,\xi^0)(z)k^{s}_{+}(z,t)\cr
           =\, &k^{e}_{-}(z,t)^{-1}L^{e}(q^0,p^0,\xi^0)(z)k^{e}_{-}(z,t).\cr}
           \eqno(5.2.12)
$$

In the following, we shall write down the solution of the 
factorization problem explicitly in terms of Riemann theta functions
for the case where $\fg = sl(N,\Bbb{C})$ with $\fh$ taken to be
the Cartan subalgebra consisting of diagonal matrices in $\fg.$
As similar procedures can also be carried out for other classical
simple Lie algebras, we shall not give details here for the other 
cases.

For our purpose, we introduce the spectral curve $C$ as defined
by the equation
$$det(L(q^0,p^0,\xi^0)(z) - wI) =0.\eqno(5.2.13)$$
By (5.2.9), this defines an $N$-sheeted branched covering 
$\pi:C\longrightarrow \Sigma$
of the elliptic curve $\Sigma.$  Let
$$I(q^0,p^0,\xi^0;z,w):=det(L(q^0,p^0,\xi^0)(z) - wI)\eqno(5.2.14)$$
for $(z,w)\in \Bbb{C}^{*}\times \Bbb{C}$.  We shall make the 
following genericity assumptions:
\smallskip
\noindent (GA1) zero is a regular value of $I(q^0,p^0,\xi^0;\cdot,\cdot),$
\newline
\noindent (GA2) the eigenvalues $\lambda_1,\ldots,\lambda_N$ of 
$\xi^0$ are distinct.
\smallskip
Then the curve $C$ is smooth.  The points on $C$ corresponding to
$z=0$ will be considered as points ``at $\infty$'', we shall denote
them by $P_1,\ldots,P_N$ respectively. 
Note that for $P$ on the finite part of $C$, we have
$dim\, ker(L(z(P)) -w(P)I) =1$, for otherwise we would obtain
a contradiction to assumption (GA1). Consequently, there exists
a unique eigenvector $\widehat{v}(P)$ of the matrix
$L^{e}(q^0,p^0,\xi^0)(z(P))$ corresponding to the
eigenvalue $w(P)$ normalized by the condition 
$\widehat{v}_1(P) = (e_1,\widehat{v}(P)) = 1$. The next
result gives a summary on the properties of the spectral
curve and $\widehat{v}(P)$ and can be obtained by
following the analysis in \c{KBBT}.

\proclaim
{Proposition 5.2.1} Under the genericity assumptions (GA1) and (GA2),
the spectral curve $C$ has the following properties:
\smallskip
\noindent (a) $C$ is smooth and is an $N$-sheeted branched cover
of the elliptic curve $\Sigma,$
\newline
\noindent (b) in a deleted neighborhood of $z=0$, $C$ can be
represented as
$$\prod_{r=1}^{N} \left({\lambda_{r}\over z} + h_{r}(z) -w\right) = 0$$
where $h_1,\ldots,h_N$ are holomorphic in a neighborhood of $z=0$,
\newline
\noindent (c) the genus of $C$ is $g ={1\over 2}(N^2 - N +2).$
\newline
On the other hand, the components 
$\widehat{v}_{j}(P) = (e_{j}, \widehat{v}(P))$ of 
the eigenvector $\widehat{v}(P)$, $j=2,\ldots,N$,
are meromorphic on the finite part of $C$ with polar
divisor $D =\sum_{i=1}^{g-1} \gamma_i.$  Moreover, in a deleted
neighborhood of $P_{k}$, we have
$$\widehat{v}_{j}(P) = e^{-\zeta(z(P))(q^{0}_j-q^{0}_1)}(\psi^{(k)}_j + O(z(P)))
\eqno(5.2.15)$$
where $\psi^{(k)}$ is the eigenvector of $\xi^0$ corresponding to
$\lambda_k$ with $\psi^{(k)}_{1} = 1$, $k =1,\ldots,N.$
\endproclaim 

Now, from the definition of $\widehat{v}(P)$ and $M^{e}(q^0,p^0,\xi^0)$, 
we have
$$M^{e}(q^0,p^0,\xi^0)(z(P))\,\widehat{v}(P) = (w(P)/z(P))\,\widehat{v}(P).
\eqno(5.2.16)$$
Hence it follows from (5.2.5) and (5.2.16) that
$$e^{t(w(P)/z(P))}(k^{s}_{+}(z(P),t)^{-1}\widehat{v}(P))=k^{e}_{-}(z(P),t)^{-1}
\widehat{v}(P)\eqno(5.2.17)$$
for $z(P)$ in a deleted neighborhood of $0\in\Sigma.$
Set
$$v_{+}(t,P) = k^{s}_{+}(z(P),t)^{-1}\widehat{v}(P),\eqno(5.2.18)$$
$$v_{-}(t,P) = k^{e}_{-}(z(P),t)^{-1}\widehat{v}(P).\eqno(5.2.19)$$
Then from (5.2.17), (5.2.12) and the definition of $\widehat{v}(P)$,
we obtain
$$e^{t(w(P)/z(P))}v_{+}(t,P) = v_{-}(t,P),\eqno(5.2.20)$$
$$L^{e}(q(t),p(t),\xi(t))(z(P))v_{\pm}(t,P) = w(P) v_{\pm}(t,P).\eqno(5.2.21)$$
In this way, we are led to scalar factorization problems for the
components of a suitably normalized eigenvector of
$L^{e}(q(t),p(t),\xi(t))(z(P))$.

\proclaim
{Proposition 5.2.2} In a deleted neighborhood of $P_k$, $k=1,\ldots,N$,
we have
$$\aligned
v^{j}_{+}(t,P)& = e^{\zeta(z(P))(q^{0}_{1}-q_{j}(t))}((k_{+}(0,t)^{-1}\psi^{(k)})_{j} + 
O(z(P)))\\
        &\sim e^{(q^{0}_{1}-q_{j}(t))z(P)^{-1}}(k_{+}(0,t)^{-1}\psi^{(k)})_{j}\quad
              as\quad P\to P_k,\\
\endaligned
$$
$$\aligned
v^{j}_{-}(t,P)& = e^{tw(P)z(P)^{-1}+\zeta(z(P))(q^{0}_{1}-q_{j}(t))}((k_{+}(0,t)^{-1}
\psi^{(k)})_{j} + O(z(P)))\\
            &\sim e^{t({\lambda_{k}/z(P)} + h_{k}(0))z(P)^{-1} +
            (q^{0}_{1}-q_{j}(t))z(P)^{-1}}
             (k_{+}(0,t)^{-1}\psi^{(k)})_{j}\quad
              as\quad P\to P_k.\\
\endaligned
$$
\endproclaim

\demo  
{Proof} This is a consequence of (5.2.18)-(5.2.20),(5.2.6),(5.2.15) and the
fact that in a deleted neighborhood of $P_k$, we have
$w(P) = \lambda_{k} z(P)^{-1} + h_{k}(0) + O(z(P))$ from
Proposition 5.2.1 (b).
\pf
\enddemo

In order to write down $v^{j}_{-}(t,P)$, we will insert a fictitious pole
together with a matching zero to this function at some point $\gamma_0$ on 
the finite part of $C$
distinct from $\gamma_1,\ldots,\gamma_{g-1}$.  By putting in an
additional pole in this way, we would be able to construct 
$v^{j}_{-}(t,P)$ as a multi-point Baker-Akheizer function.
To do so, let us fix a canonical homology basis 
$\{a_j,b_k\}_{1\leq j,k\leq g}$
of the Riemann surface associated with $C$ and let 
$\{\omega_i\}_{1\leq i\leq g}$
be a cohomology basis dual to $\{a_j,b_k\}_{1\leq j,k\leq g}$, i.e.\,\,
$\int_{a_j} \omega_i = \delta_{ij}$, $\int_{b_j} \omega_i = \Omega_{ij}$.
With respect to the Riemann matrix $\Omega = (\Omega_{ij})$, we construct
the theta function
$$\theta(z_1,\ldots,z_g) = \sum_{m\in \Bbb{Z}^g} exp\{2\pi i (m,z)
  + \pi i (\Omega m,m)\}.\eqno(5.2.22)$$
We also introduce the Abel-Jacobi map
$$A:C\longrightarrow Jac(C),\quad P\mapsto \left(\int_{P_0}^{P} \omega_1,
 \ldots, \int_{P_0}^{P} \omega_g \right)\eqno(5.2.23)$$
where $P_0$ is some fixed point on the finite part of $C$.
Now, let $d\Omega^{(i)}$, $i=1,2$, be the unique abelian differential of
second kind with vanishing $a$-periods such that in a deleted
neighborhood of $P_k$,
$$d\Omega^{(1)} = d(z^{-1} + \omega^{(1)}(z)),\eqno(5.2.24)$$
$$d\Omega^{(2)} = d(\lambda_{k} z^{-2} + h_{k}(0)z^{-1} + \omega^{(2)}(z)),
\eqno(5.2.25)$$
where $\omega^{(1)}(z), \omega^{(2)}(z)$ are regular at 
$z=0$, $k=1,\ldots,N.$  We shall
denote by $2\pi i U^{(i)}$ the vector of $b$-periods of $d\Omega^{(i)}$,
$i=1,2.$
Then from Proposition 5.2.2 and the fact that $v^{j}_{-}(t,P)$ has
$\sum_{i=0}^{g-1} \gamma_i$ as a polar divisor, we obtain the following
result from the standard construction of Baker-Akhiezer functions \c{K}.

\proclaim
{Proposition 5.2.3} For $1\leq j\leq N$, 
$$\aligned
   &v^{j}_{-}(t,P)\\
=\,&f_{j}(t)\frac{\theta(A(P)+(q^{0}_{1}-q_{j}(t))U^{(1)}+tU^{(2)}-A(D)-A(\gamma_0)
    -K)}{\theta(A(P)-A(D)-A(\gamma_0)-K)}\\
   &\times exp\,[(q^{0}_{1}-q_{j}(t))\Omega^{(1)}(P)+t\,\Omega^{(2)}(P)]\\
\endaligned
$$
where 
$$\Omega^{(i)}(P) =\int_{P_0}^{P} d\,\Omega^{(i)},\quad i=1,2$$
and $K$ is the vector of Riemann constants.
\endproclaim

\proclaim
{Corollary 5.2.4} $\theta\left(q_{j}(t)U^{(1)}-tU^{(2)} +V\right) =0$ where
$V =A(D)-q^{0}_{1}U^{(1)}+K,\, j=1,\ldots,N.$
\endproclaim

\demo
{Proof} This follows when we evaluate the expression for 
$v^{j}_{-}(t,P)$ in Proposition 5.2.3 at the point $\gamma_0$
and equate the result to zero.
\pf
\enddemo

Let
$$f(t) = diag (f_1(t),\ldots,f_N (t)).\eqno(5.2.26)$$
In view of Proposition 5.2.3, we shall write
$$v_{-}(t,P) = f(t)\,v^{\theta}_{-}(t,P)\eqno(5.2.27)$$
where the $v^{\theta}_{-}(t,P)$ are known.  Note that if we
set $t=0$ in the above expression, we obtain 
$\widehat{v}(P) = f(0)\,v^{\theta}_{-}(0,P).$
Clearly, $f_{1}(0)=1$; the other $f_{j}(0)$'s are then
uniquely determined from the definition of $\widehat{v}(P)$.
Now, for given $z\in\Sigma$ which is not a branch point of the
coordinate function $z(P)$, there exist $N$ points
$P_{1}(z),\ldots,P_{N}(z)$ of $C$ lying over $z$.  Hence
we can define the matrices
$$\eqalign{\widehat{V}(z)=(\widehat{v}(P_{1}(z)),&\ldots,\widehat{v}(P_{N}(z))),
  \,\,\,V_{-}(z,t) =(v_{-}(t,P_{1}(z)),\ldots,v_{-}(t,P_{N})),\cr
 &V^{\theta}_{-}(z,t) = (v^{\theta}_{-}(t,P_{1}(z)),\ldots,
  v^{\theta}_{-}(t,P_{N}(z))).\cr}\eqno(5.2.28)$$
With these definitions, if follows from  (5.2.27) and (5.2.19) that
$$k^{e}_{-}(z,t) = \widehat{V}(z) V^{\theta}_{-}(z,t)^{-1}f(t)^{-1}
\eqno(5.2.29)$$
where $f(t)$ is still to be determined.  To do this, we invoke
the condition that 
$T_{\gamma_{+} (t)} {\bbl}_{{\gamma_{+}(t)}^{-1}} {\dot\gamma_{+}(t)}
\in {\Cal R}^{+}(\lbrace q(t) \rbrace \times \lbrace 0 \rbrace \times L\fg).$
Indeed, it follows from the proof of Theorem 2.2.2 in \c{L2} that
$$T_{\gamma_{+} (t)} {\bbl}_{{\gamma_{+}(t)}^{-1}} {\dot\gamma_{+}(t)}
= {\Cal R}^{+}(q(t),0, M(q(t),p(t),\xi(t))).\eqno(5.2.30)$$
Consequently, we have
$$\eqalign{& k_{+}(z,t)^{-1}\dot{k}_{+}(z,t)\cr
  =\,& (R^{+}(q(t))M(q(t),p(t),\xi(t)))(z)\cr
  =\,&  M(q(t),p(t),\xi(t))(z) -\zeta(z)p(t) +\sum_{i\neq j}
             l(q_{i}(t)-q_{j}(t),z)(\zeta(q_{i}(t)-q_{j}(t))\cr
        \quad & +\zeta(z)-
             \zeta(q_{i}(t)-q_{j}(t)+z))
             \xi_{ij}(t) e_{ij}\cr}\eqno(5.2.31)$$
where in the last step we have used (5.1.8).  Therefore, when
we expand the above expression about $z=0$ and compare the
term in $z^{0}$, we find that 
$$k_{+}(0,t)^{-1}\dot{k}_{+}(0,t) =\sum_{i\neq j} \zeta^{\prime}
  (q_{i}(t)-q_{j}(t))\xi_{ij}(t) e_{ij}\eqno(5.2.32)$$
from which we obtain the condition
$$\Pi_{\fh} \left((k_{+}(0,t)^{-1}\dot{k}_{+}(0,t)\right)=0.\eqno(5.2.33)$$
In order to state our next result, we introduce 
$$\eqalign{&
\omega^{(1)}_{k} = \lim_{P\to P_k} \left(\Omega^{(1)}(P) -z(P)^{-1}\right)\cr
&\omega^{(2)}_{k} = \lim_{P\to P_k} \left(\Omega^{(2)}(P)-(\lambda_{k} z(P)^{-2}+
  h_{k}(0)z(P)^{-1})\right)\cr}\eqno(5.2.34)$$
for $k=1,\ldots,N$.  We also introduce the matrix 
$W^{\theta}(t)=(W^{\theta}_{jk}(t))$
where
$$\eqalign{W^{\theta}_{jk}(t)&
=\, \frac{\theta(A(P_k)+(q^{0}_{1}-q_{j}(t))U^{(1)}+tU^{(2)}-A(D)-A(\gamma_0)
    -K)}{\theta(A(P_k)-A(D)-A(\gamma_0)-K)}\cr
   &\quad \times exp\,\left[(q^{0}_{1}-q_{j}(t))\omega^{(1)}_{k}+t\omega^{(2)}_{k}
    \right].\cr}
   \eqno(5.2.35)$$

\proclaim
{Proposition 5.2.5} $f(t)$ satisfies the differential equation
$$\dot{f}(t) = \, -f(t)\,\Pi_{\fh}\left(\dot{W}^{\theta}(t)W^{\theta}(t)^{-1}\right)$$
and hence
$$f(t) = f(0)\,exp\left\{-\int_{0}^{t} \Pi_{\fh}\left(\dot{W}^{\theta}(\tau)
  W^{\theta}(\tau)^{-1}\right)\,d\tau\right\}.\eqno(5.2.36)$$
\endproclaim

\demo
{Proof} From Proposition 5.2.2 and 5.2.3, we have
$$\aligned
         &(k_{+}(0,t)^{-1}\psi^{(k)})_{j}\\
     =\,&\lim_{P\to P_k} v^{j}_{-}(t,P) exp\,[-t(\lambda_{k}z(P)^{-2}+
          h_{k}(0)z(P)^{-1})-(q^{0}_{1}-q_{j}(t))z(P)^{-1}]\\
     =\, &\, f_{j}(t) W^{\theta}_{jk}(t)\\
\endaligned
$$
which implies
$$k_{+}(0,t) = \Psi W^{\theta}(t)^{-1} f(t)^{-1}$$
where $\Psi$ is the $N\times N$ matrix whose $k$-th column is the vector 
$\psi^{(k)}$, $k=1,\ldots, N.$
Differentiating the above expression with respect to $t$, we find
$$k_{+}(0,t)^{-1}\dot{k}_{+}(0,t) = -f(t)\dot{W}^{\theta}W^{\theta}(t)^{-1} f(t)^{-1}
-\dot{f}(t)f(t)^{-1}.$$
Therefore, when we apply the condition in (5.2.33), the desired equation for 
$f(t)$ follows.  Finally, the solution of the equation is obvious.
\pf
\enddemo

Hence $k_{+}(0,t) = \Psi W^{\theta}(t)^{-1} f(t)^{-1}$ and
$k^{e}_{-}(z,t) = \widehat{V}(z) V^{\theta}_{-}(z,t)^{-1}f(t)^{-1}$
are completely determined.  Therefore we have following
result.

\proclaim
{Theorem 5.2.6} Let 
$(q^0,p^0,\xi^0)\in J^{-1}(0) = TU\times ({\Cal U}\cap \fh^{\perp})$
satisfy the genericity assumptions (GA1), (GA2).  Then  the
Hamiltonian flow on $J^{-1}(0)$ generated by
$${\Cal H}(q,p,\xi)  = {1\over 2} \sum_{i} p_{i}^{2} - {1\over 2}
  \sum_{i\neq j} \wp(q_i-q_j) {\xi_{ij}\xi_{ji}}$$
with initial condition $(q(0),p(0),\xi(0)) =(q^0,p^0,\xi^0)$
is given by
$$\eqalign{
         &\theta\left(q_{j}(t)U^{(1)}-tU^{(2)}+V\right)=0,\quad j=1,\ldots,N,\cr
         &\xi(t)= k_{+}(0,t)^{-1}\xi^{0}\,k_{+}(0,t),\cr
         &p(t)=Ad_{k^{e}_{-}(z,t)^{-1}} L^{e}(q^0,p^0,\xi^0)(z)\cr
         &\,\,\,\quad\,\,\, +\sum_{i\neq j} l(q_i(t)-q_j(t),z)e^{-\zeta(z)(q_i(t)-q_j(t))}
          \xi_{ij}(t) e_{ij}\cr}\eqno(5.2.37)$$
where $k_{+}(0,t)$, $k^{e}_{-}(z,t)$ are given by the formulas
above.
\endproclaim

Finally we are ready to give the solutions of the associated 
integrable model on $TU\times \fg_{red}$ whose equations
are given in Proposition 5.1.2.

\proclaim
{Corollary 5.2.7} Let $(q^0,p^0,s^0) \in TU \times \fg_{red}$ and
suppose $s^0 =  Ad_{g(\xi^{0})^{-1}} \xi^{0}$ where 
$\xi^{0}\in {\Cal U} \cap \fh^{\perp}.$  Then the Hamiltonian
flow generated by ${\Cal H}_{0}$ with initial condition
$(q(0),p(0),s(0)) = (q^0,p^0,s^0)$ is given by
$$\eqalign{
 &\theta\left(q_{j}(t)U^{(1)}-tU^{(2)}+V\right)=0,\quad j=1,\ldots,N,\cr
         &s(t)= \, Ad_{\bigl({\widetilde k}_{+}(0,t)\,
                  g \bigl(Ad_{{\widetilde k}_{+}(0,t)^{-1}}\, s^{o}\bigr)
                \bigr)^{-1}} s^{0},\cr
         &p(t)=Ad_{\bigl({\widetilde k}^{e}_{+}(z,t)\,
                  g \bigl(Ad_{{\widetilde k}_{+}(0,t)^{-1}}\,s^{0}\bigr)
                  \bigr)^{-1}} L^{e}(q^0,p^0,s^0)(z)\cr
         &\,\,\,\quad\,\,\, +\sum_{i\neq j} l(q_i(t)-q_j(t),z)e^{-\zeta(z)
            (q_i(t)-q_j(t))}s_{ij}(t) e_{ij}\cr}\eqno(5.2.38)$$
where $k_{+}(0,t)$, $k^{e}_{-}(z,t)$ are given by the formulas
above.
\endproclaim

\noindent {\bf Remark 5.2.8.} (a) In \c{KBBT}, the authors
considered the $gl(N,\Bbb C)$-elliptic spin Calogero-Moser system
with Hamiltonian
$$H(q,p,f) = {1\over 2} \sum_{i=1}^{N} p_{i}^{2} + {1\over 2}
  \sum_{i\neq j} \wp(q_i-q_j) {\xi_{ij}\xi_{ji}} \eqno(5.2.39)$$
and they imposed the following restriction on 
$\xi = (\xi_{ij})\in  gl(N,\Bbb C)\simeq gl(N,\Bbb C)^{*}$, 
namely, they set 
$$\xi_{ij} = b^{T}_{i}a_{j}\eqno(5.2.40)$$
for all $i$ and $j$ where
$a_{j}$, $b_{j}$ 
are (column) vectors in $\Bbb C^{l},$
$l<N,$ satisfying the nontrivial Poisson bracket relations
$\{a_{i,\alpha}, b_{j,\beta}\} =\delta_{i,j}\delta_{\alpha,\beta}.$ 
Thus from the outset, it is clear that these authors 
were restricting themselves to special
coadjoint orbits of $gl(N,\Bbb C)^{*}\simeq gl(N,\Bbb C)$
which consist of matrices of the form
$B^{T}A$, where
$$A =(a_1,\cdots, a_N),\,\,\,B=(b_1,\cdots, b_N)\eqno(5.2.41)$$
are $l\times N$ matrices.  However, it is only through the
imposition of (5.2.40) that they were able
to make the connection with the matrix KP
equation.  The precise relation is that the equations of motion for
$a_j$, $b_j$ (up to gauge equivalence) and $q_{j}$
are the necessary and sufficient condition for
the time-dependent matrix Schr\"odinger equation
$$\left(\partial_{t} -\partial^{2}_{x} + \sum_{j=1}^{N} a_{j}(t)b^{T}_{j}(t)
\wp(x-q_{j}(t))\right)\Psi =0\eqno(5.2.42)$$
and its adjoint
$${\widetilde\Psi}^{T}\left(\partial_{t} -\partial^{2}_{x} + \sum_{j=1}^{N} 
a_{j}(t)b^{T}_{j}(t)\wp(x-q_{j}(t))\right)\ =0\eqno(5.2.43)$$
(${\widetilde\Psi}^{T}\partial \equiv -\partial {\widetilde\Psi}^{T}$)
to admit solutions of the form
$$\Psi=\sum_{j=1}^{N} s_{j}(t,k,z)\Phi(x-q_{j}(t),z)e^{kx + k^{2}t},\eqno(5.2.44)$$
$${\widetilde\Psi}=\sum_{j=1}^{N} s^{+}_{j}(t,k,z)\Phi(-x+q_{j}(t),z)
e^{-kx- k^{2}t},\eqno(5.2.45)$$
where $s_{j}$,$s^{+}_{j}$ are functions which take values in $\Bbb C^{l}$
and 
$$\Phi(x,z) =\frac{\sigma(z-x)}{\sigma(x)\sigma(z)}e^{\zeta(z)\,x}.\eqno(5.2.46)$$
Furthermore, it is in the course of proving this result that the Lax pair
as well as the constraint $f_{jj} = b^{T}_{j} a_{j} =2, j=1,\cdots, N,$ emerge
naturally.  In short, the method of solution in \c{KBBT} of their
elliptic spin Calogero-Moser system as defined in
(5.2.39), (5.2.40) on the constraint manifold $\{f_{jj}= b^{T}_{j}a_{j}=2,
j=1,\cdots,N\}$ is based 
on the above correspondence. 

In what follows, we shall give a sketch of this method
to solve the equations of motion for $a_{j}(t), b_{j}(t)$ and $q_{j}(t)$
so that the reader can understand its limitations.
To start with, one has a normalized Baker-Akhiezer vector function
$\psi(x,t,P)$ (and its dual $\psi^{+}(x,t,P)$) which is uniquely 
determined by the spectral curve, a divisor
of degree $g+l-1$, and prescribed behaviour in deleted
neighborhoods of the punctures $P_j$, $j=1,\cdots,l.$ 
(These are in turn fixed by the initial data.)
From a general result in \c{K}, corresponding to 
$\psi(x,t,P)$ and $\psi^{+}(x,t,P)$ is an algebro-geometric potential
$u(x,t)$ satisfying
$$\left(\partial_{t} -\partial^{2}_{x} +  u(x,t)\right)
\psi(x,t,P) =0\eqno(5.2.47)$$
and 
$$(\psi^{+}(x,t,P))^{T}\left(\partial_{t} -\partial^{2}_{x} + u(x,t)\right)\ =0.
\eqno(5.2.48)$$
Next, by analyzing $\psi(x,t,P)$ as a function of $x$, and
on comparing $\Psi(x,t,P)$ with $\psi(x,t,P)$, one
can show that there exists a constant invertible matrix $\chi_{0}$ such that
$$\chi_{0} u(x,t) \chi^{-1}_{0} = \sum_{j=1}^{N} a_{j}(t)b^{T}_{j}(t)
\wp(x-q_{j}(t)).\eqno(5.2.49)$$
Since $u(x,t)$ is determined by $\psi(x,t,P)$, it is in this fashion that 
the authors in \c{KBBT} were
able to write down $a_j(t), b_j(t)$ and the equation satisfied
by $q_j(t)$ in terms of theta
functions.  Now, let us examine the expression
$\sum_{j=1}^{N} a_{j}(t)b^{T}_{j}(t)\wp(x-q_{j}(t))$
for the potential carefully.  Clearly, it depends on the $a_j$'s  and $b_j$'s
through the rank one matrices $a_{j}b^{T}_{j}$
rather than on the entries of the matrix $\xi = (b_{j}^{T} a_{j}).$  
For this 
reason, the solution method sketched above is rather
specific to these special coadjoint orbits of $gl(N,\Bbb C)^*.$
Obviously, similar remarks also hold for the corresponding
rational and trigonometric cases.
Thus it is clear that this method is not applicable to our 
more general class of spin Calogero-Moser systems associated with
simple Lie algebras here.
\smallskip
\noindent (b) The first
link between elliptic solutions of integrable PDEs and discrete
particle systems was found in the paper of Airault, McKean and
Moser \c{AMM}.  Indeed, the PDE in \c{AMM} is KdV and the 
corresponding discrete particle system is the usual (spinless) 
Calogero-Moser system.  In the context of the spin CM system
defined by (5.2.39) and (5.2.40) above, we remark that
its correspondence with matrix KP is related to some interesting
algebraic geometry and we refer the reader to \c{T}
for details.  For our general class of spin CM
systems which we address in this work, whether
it has any connection with integrable PDEs is
entirely open at this point.

\bigskip
\noindent{\bf Acknowledgments.} The factorization method in \c{L2}
and the present work has been the subject of several lectures in the last 
few years. The author would like to express his appreciation to 
his colleagues at various institutions
for their interest and for their hospitality during his visits.
He would also like to thank Alan Weinstein for an inspiring
lecture at MSRI in 1989 which set him to learn about groupoids. 
Special thanks is due to Armando Treibich for an interesting
explanation on tangential covers at IHP in the
summer of 2004. Finally, he acknowledges the helpful advice of Percy Deift
and an anonymous referee on the presentation of the material.

\enddocument